\documentclass{article}
\usepackage{graphicx}

\title{Quantum-enabled Spintronic ``Small'' Antennas}

\author{Supriyo Bandyopadhyay\\
Department of Electrical and Computer Engineering \\ Virginia Commonwealth University, Richmond, VA 23284, USA \\
Email: sbandy@vcu.edu}

\begin{document}

\maketitle

\begin{abstract}
Antennas transmit information wirelessly from one location to another via electromagnetic waves. Miniaturizing them, however, is challenging since the radiation efficiencies of all traditional antennas, based on the principles of classical electromagnetics, plummet when their dimensions are shrunk to tiny fractions of the radiated wavelength. Lately, a new generation of antennas whose operations are underpinned by non-classical principles have been demonstrated and they can overcome this limitation. This enables embedded applications that were hitherto inaccessible. In addition, beam steering, which normally requires a large phased array (multiple antenna elements each much larger than the wavelength), can now be accomplished with a single element much smaller than the wavelength. Some of these antennas also have stealth attributes for secure and covert communication, which makes this new genre a disruptive new technology.

\end{abstract}

\section{Introduction}
Small antennas that are sub-wavelength in size are known to be inefficient radiators. The radiation efficiency of an antenna is given by the expression $\eta = R_{rad}/\left ( R_{rad} + R_{loss} \right )$, where $R_{rad}$ is the radiation resistance and $R_{loss}$ is the resistance associated with losses in the antenna. The radiation resistance of a short dipole antenna, for example, is given by $R_{rad} = 20 \pi^2 \left (A/\lambda \right )^2$ ohms \cite{illinos}, which immediately suggests that the radiation efficiency varies as $\left (A/\lambda \right )^2$. Therefore, the efficiency will suffer if the antenna dimension is much smaller than the wavelength, i.e., if $A \ll \lambda^2$. This was more formally established by a number of authors \cite{Harrington,skrivervik,li1} and came to be known as the ``Harrington limit'' which stipulated that the maximum radiation efficiency of a small (sub-wavelength) antenna operating on classical electromagnetic principles  will be limited to $\sim$ $\left (A/\lambda \right )^2$. This has hindered antenna miniaturization for the last several decades. 

\begin{figure}[!h]
\centering
\includegraphics[width=0.99\textwidth]{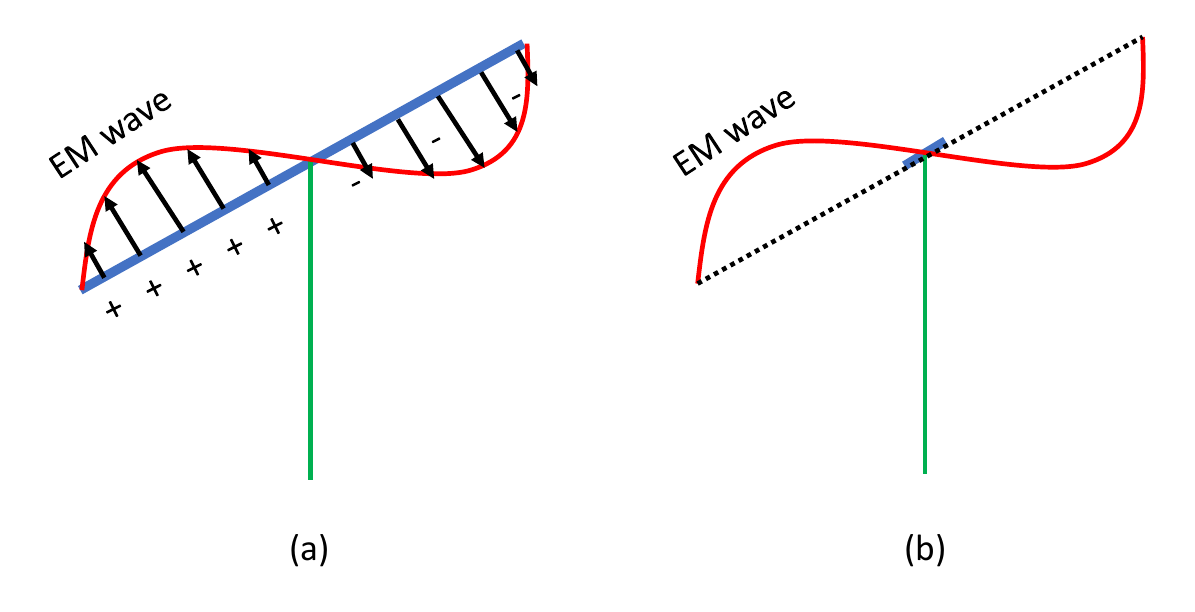}
\caption{(a) A dipole antenna that is one wavelength long and can host a resonant electric dipole. (b) An antenna much smaller than the wavelength cannot host a dipole.}
\label{fig:dipole}
\end{figure}

A qualitative understanding of the Harrington limit can be gained from Fig. \ref{fig:dipole}. Classical electromagnetic antennas radiate via fluctuating charges or fluctuating electric dipoles. If the antenna's dimension is much smaller than the wavelength, then electric dipoles cannot form in the antenna, as shown in Fig. \ref{fig:dipole}(b), and hence the radiation, if present at all, will be very weak. If the antenna is made of a magnetic material, then fluctuating magnetic dipoles can also form in them, which radiate but much less effectively than fluctuating electric dipoles. Thus, classical antennas are unable to radiate efficiently when they are made much smaller than the wavelength of radiation. This has been the understanding for the last many decades and has stymied antenna miniaturization.
\begin{figure}[!h]
    \centering
    \includegraphics[width=0.91\linewidth]{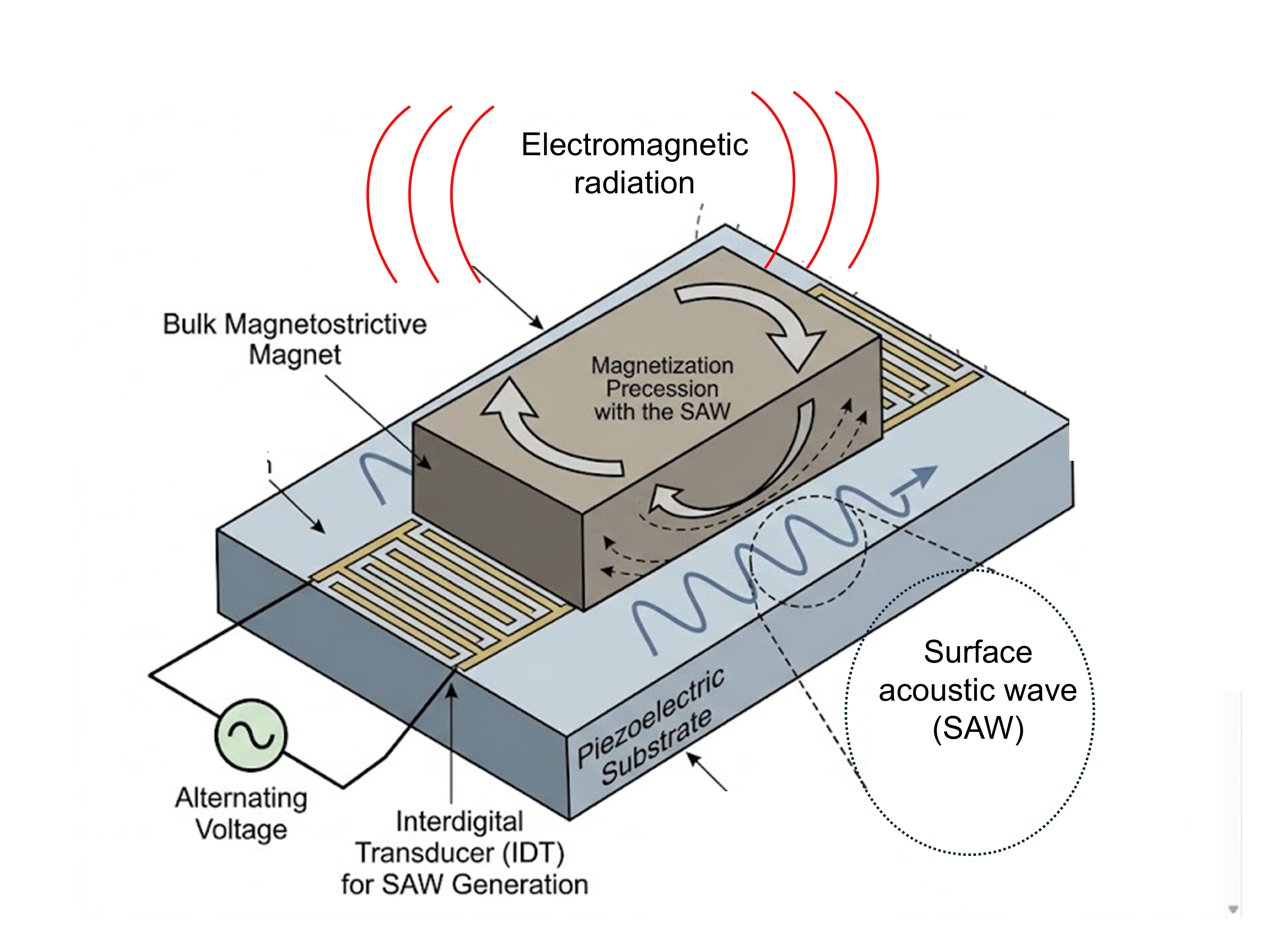}
    \caption{A bulk {\it magnetostrictive} magnet on a piezoelectic substrate or film will radiate electromagnetric waves if a surface acoustic wave (SAW) [or a bulk acoustic wave] is launched in the piezoelectric. The acoustic wave generates time varying strain in the magnet and that causes the magnetization to precess because of the inverse magnetostriction (or Villari) effect. The time-varying magnetization radiates electromagnetic waves and this is the basis of a magneto-elastic antenna. }
        \label{fig:bulk}
\end{figure}

\section{Classical magneto-elastic antennas}

The first attempt to overcome this limitation involved magneto-elastic (or acoustic) antennas \cite{carman,sun,liu1,liu3,luo,ma,xu,zhang,fu2,hu,schneider,sarkar}. In these constructs, a bulk {\it magnetostrictive} magnet is deposited on a piezoelectric substrate and a surface or bulk acoustic wave is launched in the piezoelectric. The acoustic wave subjects the magnet to periodic strain and causes its magnetization to precess with the frequency of the acoustic wave (i.e., the periodic strain) because of the {\it inverse magnetostriction} (or Villari) effect. This precessing magnetization radiates electromagnetic waves in space, thereby implementing an antenna, as shown in Fig. \ref{fig:bulk}.

These antennas can have much higher radiation efficiencies than what the Harrington limit of $\sim\left (A/\lambda \right )^2$ will allow because they are driven by {\it acoustic resonances} instead of the usual electromagnetic resonance. That will make the wavelength $\lambda$ in the Harrington expression become the acoustic (or SAW) wavelength which is typically five orders of magnitude smaller than the electromagnetic wavelength for the same frequency in most piezoelectrics. Consequently, the Harrington limit increases by approximately ten orders of magnitude. However, there are other effects that come into play and severely reduce the radiation efficiency, such as eddy current losses in the bulk magnets and Gilbert damping which causes the magnetization precession to decay with time. As a result, the efficiency can exceed $A/\lambda_{EM}^2$, but still falls far short of $A/\lambda_{ac}^2$ by several orders of magnitude, where $\lambda_{EM}$ is the electromagnetic wavelength and $\lambda_{ac}$ is the acoustic wavelength at the frequency of radiation.

\nopagebreak

\subsection{Replacing bulk magnets with nanomagnets}

\begin{figure}[!t]
    \centering
    \includegraphics[width=0.91\linewidth]{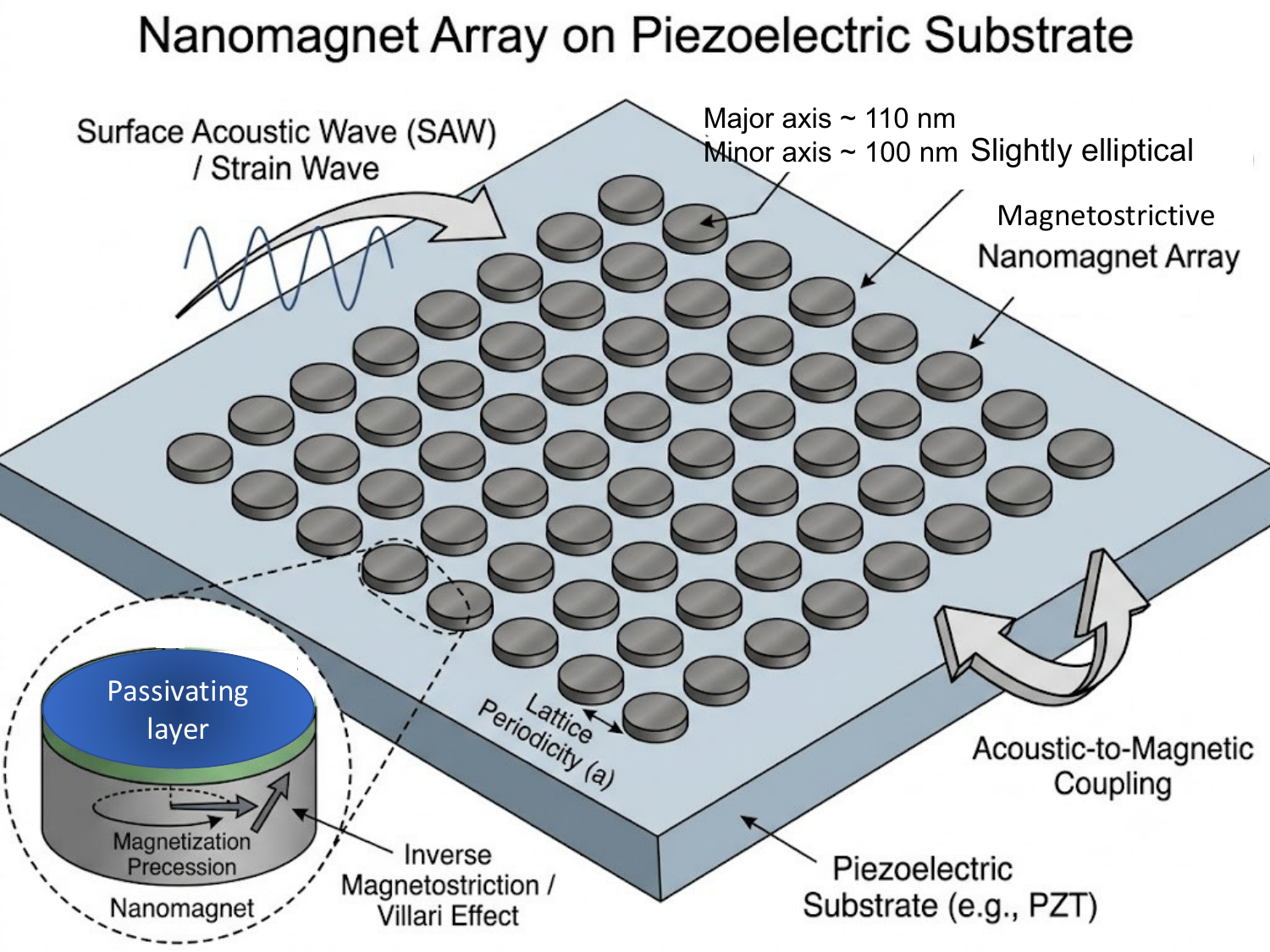}
    \caption{A two-dimensional periodic array of slightly elliptical magnetostrictive nanomagnets deposited on a piezoelectric substrate (or a piezoelectric thin film) forming an ``artificial multiferroic magnonic crystal''. A surface acoustic wave launched in the piezoelectric with appropriate electrodes (not shown) will subject the nanomagnets to periodic strain and cause their magnetizations to precess owing to the inverse magnetostriction effect. This will  radiate electromagnetic waves in the surrounding space. }
    \label{fig:array}
\end{figure}

The radiation efficiency of magneto-elastic antennas can be improved dramatically by replacing bulk magnets with a two-dimensional periodic array of  ``nanomagnets'' which suppress the formation of eddy current loops (because of their small size) and hence reduce eddy current losses. This greatly reduces $R_{loss}$ and increases the radiation efficiency $\eta = R_{rad}/\left ( R_{rad} + R_{loss} \right )$. Fig. \ref{fig:array} shows a schematic of this antenna.

\begin{figure}[!h]
    \centering
    \includegraphics[width=0.99\linewidth]{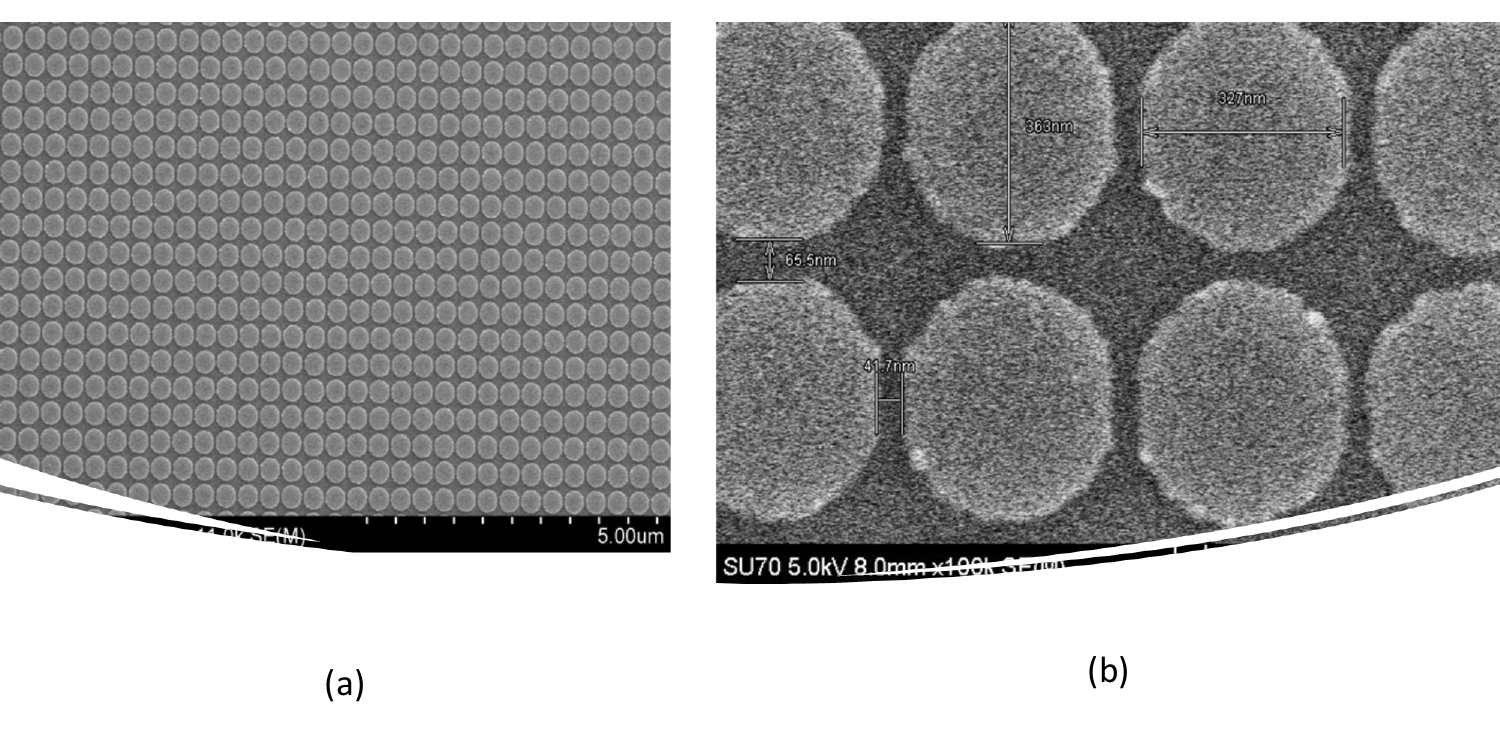}
    \caption{(a) Scanning electron micrograph of a two-dimensional periodic array of nanomagnets deposited on a piezoelectric LiNbO$_3$ substrate in which a RF frequency SAW was launched with electrodes. (b) Higher resolution image showing the nanomagnet dimension and inter-nanomagnet spacing. The nanomagnets are elliptical with major axis dimension $\sim$360 nm, minor axis $\sim$330 nm, edge-to-edge spacing $\sim$65 nm along the major axis direction and $\sim$42 nm along the minor axis direction. The thickness of the nanomagnets was 6 nm. Reproduced from \cite{drobitch} with CC-BY license.}
    \label{fig:SEM}
\end{figure}

The nanomagnet-approach to a magneto-elastic antenna was first introduced in \cite{drobitch} and later also adopted in \cite{fu1}. Fig. \ref{fig:SEM} shows a scanning electron micrograph of the two-dimensional periodic array of nanomagnets (made of magnetostrictive cobalt) that was used in \cite{drobitch}. The array was deposited on a LiNbO$_3$ substrate in which a RF frequency surface acoustic wave (SAW) was launched with electrodes. The SAW caused magnetization precession in the nanomagnets via the Villari effect (this time unfettered by eddy current losses) and hence relatively strong emission of electromagnetic waves. The electromagnetic emission was measured with a dipole antenna which was placed more than four electromagnetic wavelengths away from the sample to ensure that only the far-field radiation is measured. The detector was calibrated for 144 MHz and 900 MHz, and hence these two frequencies were used to launch the SAW and measure the emitted far-field radiation.

\begin{figure}[!hbt]
\centering
\includegraphics[width=0.99\textwidth]{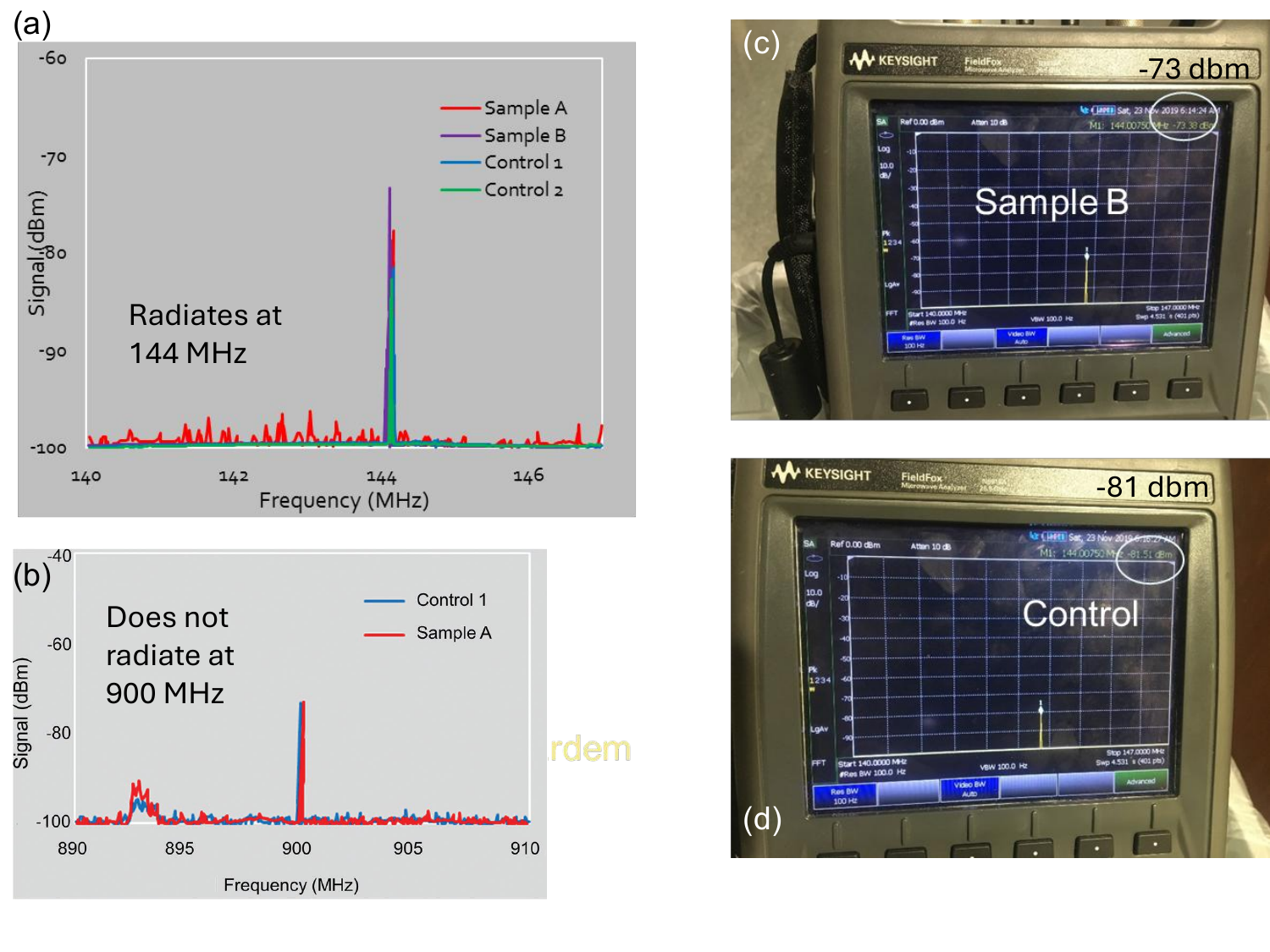}
\caption{Electromagnetic radiation spectrum of the real and control samples at 144 MHz and 900 MHz. Two real samples $A$ and $B$ were measured along with their corresponding control samples that contained no nanomagnets. (a) Radiation spectra of the two samples $A$ and $B$ along with those of the corresponding control samples at 144 MHz input SAW frequency. (b) The radiation from a real sample and the corresponding control sample cannot be resolved at 900 MHz input SAW frequency indicating that the nanomagnets are not radiating sufficiently. (c) Direct screenshot from the spectrum analyzer for sample B at 144 MHz SAW frequency. The received power at the detector is -73 dbm. (d) Direct screenshot from the spectrum analyzer for the corresponding control sample at 144 MHz SAW frequency. The received power in this case is -81 dbm at the detector. This difference of 8 db strongly suggests that the nanomagnets are radiating. Reproduced from \cite{drobitch} with CC-BY license.}
\label{fig:spectrum}
\end{figure}

Electromagnetic emission was detected when the exciting SAW frequency was 144 MHz, but not when it was 900 MHz. At the higher frequency, large angle magnetization precession could not occur probably because the magnetization could not keep pace with the rapidly varying strain generated by the high-frequency SAW. The detected emission was frequency-resolved with a spectrum analyzer and the measured spectra are shown in Fig. \ref{fig:spectrum} for both 144 MHz and 900 MHz SAW frequency. 

In order to eliminate any spurious emission that could occur from these samples owing to radiation by the SAW-launching-electrodes, cables, etc., two nominally identical samples were tested at each frequency, with the difference being that one contained nanomagnets and the other did not. The former was termed the ``real sample'' and the latter the ``control sample''. The emission spectra of both are shown in Fig. \ref{fig:spectrum}. For the same input SAW power, the radiation from the real sample exceeded that from the control sampleby 8 db (or 6.3$\times$) at 144 MHz. This was a large enough difference to assert with some confidence that the nanomagnets are radiating electromagnetic waves, although the exact radiated power cannot be determined accurately from this measurement. 

Multiple real and corresponding control samples were tested. Sample A in \cite{drobitch} had 55,000 nanomagnets and an area of 5$\times$10$^{-9}$ m$^2$ while sample B had 275,000 nanomagnets and an area of 2.5$\times$10$^{-8}$ m$^2$. The nanomagnets were slightly elliptical in shape with major axis $\sim$360 nm and minor axis $\sim$330 nm. At 144 MHz frequency, the ratio $A/\lambda_{EM}^2$ for sample A was 1.25$\times$10$^{-9}$ and for sample B it was 6.25$\times$10$^{-9}$. Therefore, these are extreme sub-wavelength antennas that would have had minuscule radiation efficiencies had they been constrained by the Harrington limit. However, based on the measurements made, the estimated radiation efficiency in sample A was 1.4$\times$10$^{-4}$ and in sample B it was 8.8$\times$10$^{-4}$ \cite{drobitch}. These exceeded the $A/\lambda_{EM}^2$ ratio by 1.2$\times$10$^5$ and 1.4$\times$10$^5$, respectively, i.e.,  by more than {\it five orders of magnitude} in both samples. This not only upended the Harrington limit that had stood for more than six decades, but did so overwhelmingly.

One possible application area for such an RF frequency nanomagnetic antenna is RFID
which uses 125–134 kHz for low frequency and 13.56 MHz
for high frequency in North America. Because of the low
power output, these  antennas will be
usable for short-range devices like access control (door and
gate openers), alarms, medical implants, metering devices,
remote control, and telemetry -- to name a few. Medically implanted devices (brain implants, pace makers, etc.) \cite{pournoori} are especially suitable because they do
not prefer high frequencies above 1 GHz to avoid tissue damage. The need
is for a low-frequency (long wavelength) antenna that is much
smaller than the wavelength (so it can be minimally invasive) and yet
can radiate with acceptable efficiency. These nanomagnet-based magneto-elastic antennas may be able to fulfill that need. Furthermore, the piezoelectric component may serve a dual role; it could be useful for harvesting energy from the patient's body movements alone and that would eliminate the need for a battery. In turn, that avoids unnecessary surgeries to just replace the battery.
Ultimately, these antennas enable aggressive miniaturization which benefits many embedded applications.

\section{Antennas based on tripartite phonon-magnon-photon coupling in artificial multiferroic magnonic crystals (AMMC)}

\begin{figure}[!h]
    \centering
    \includegraphics[width=0.99\linewidth]{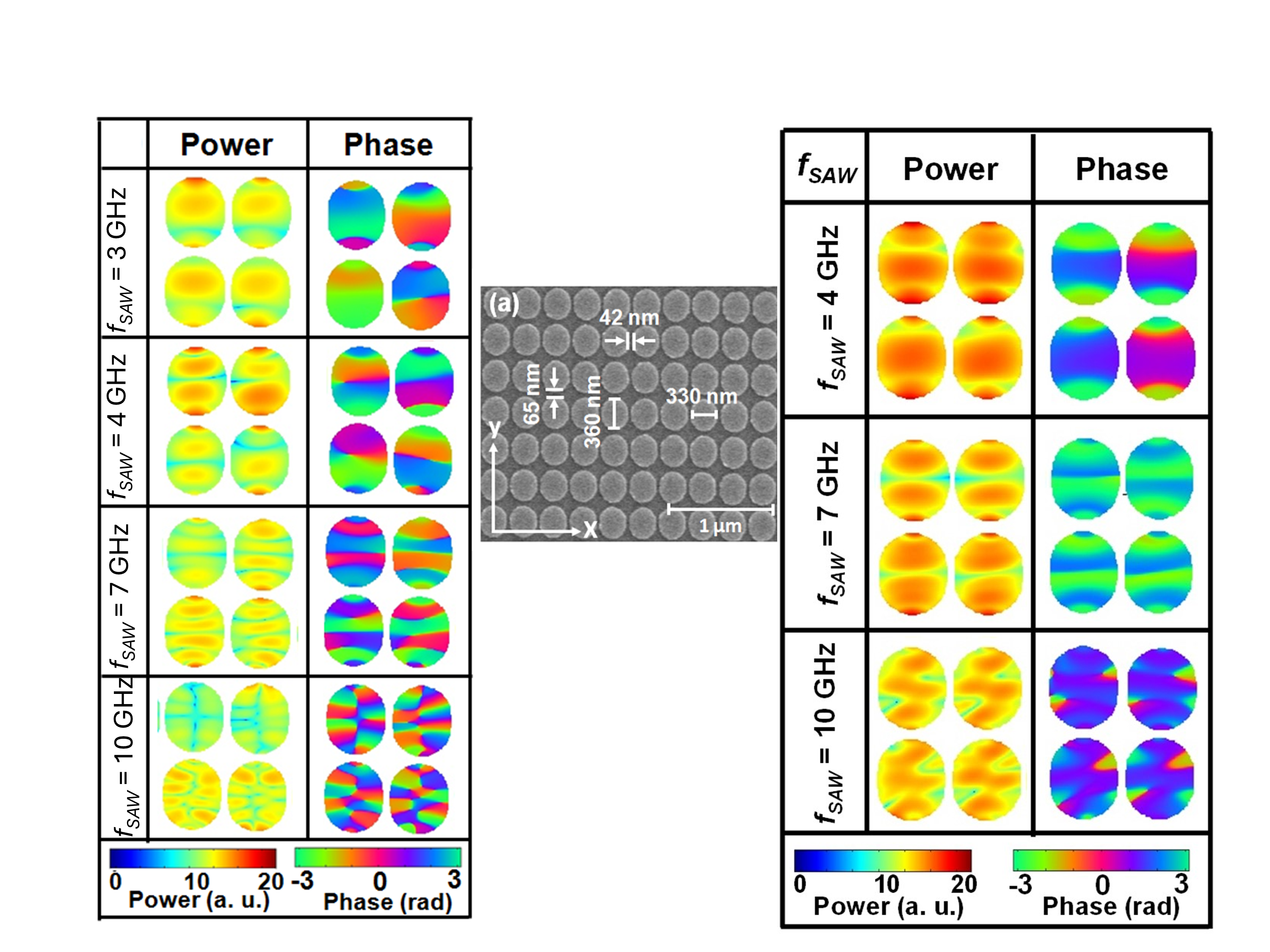}
    \caption{The simulated power and phase profiles of the intrinsic hybrid magneto-dynamical modes in a 4-nanomagnet array on a piezoelectric substrate. The nanomagnets are made of magnetostrictive cobalt. There are four intrinsic modes when the SAW propagates parallel to the major axes of the nanomagnets as determined by time-resolved magneto-optical Kerr effect microscopy (left panel) and there are three intrinsic modes when the SAW propagates parallel to the minor axes of the nanomagnets (right panel). The profiles are not the same in all four nanomagnets at  any frequency because of inter-nanomagnet magnetostatic coupling. The scanning electron micrograph of the array is shown in the center. It does not have rotational symmetry in space because the nanomagnets are elliptical (major axis dimension $\neq$ minor axis dimension) and the edge-to-edge spacing is also different in the two directions. All modes are quantized modes. Adapted from \cite{nanoscale} with permission of the Royal Society of Chemistry.}
    \label{fig:profile}
\end{figure}

When a SAW is launched into a piezoelectric film or substrate overlaid with a periodic array of nanomagnets, the magnetization precession that results within the nanomagnets from magneto-elastic coupling with the time-varying strain generated by the SAW can excite specific types of spin wave modes in the nanomagnets. These  are not pure spin waves but rather {\it hybrid magneto-dynamical modes} \cite{sucheta} that have a mixed spin- and acoustic-wave character since they are born of coupling between a phonon in the SAW and a magnon in the spin wave. These hybrid modes have rich power and phase textures as shown in Fig. \ref{fig:profile}. 

The size, shape, material and geometry of the nanomagnets, as well as the array pitch, determine the so-called ``intrinsic'' hybrid magneto-dynamical modes of the array, which are characteristic of the array. If the array does not have rotational symmetry is space, then the direction of SAW propagation also affects the intrinsic modes. Each intrinsic mode has its own frequency, as well as power and phase profile within a nanomagnet, and these quantities are determined by the size, shape, geometry and array pitch of the nanomagnets.

When the SAW frequency {\it matches} an intrinsic mode frequency, the excitation becomes ``resonant'' and very efficient transfer of power from the SAW to the intrinsic mode can take place. When it does not match, the excitation is non-resonant and in that case the SAW  will excite a mode at its own frequency (which we call an {\it extrinsic} mode) {\it plus} one or more of the intrinsic modes which are at different frequencies than the SAW frequency \cite{nanoscale}, provided the intrinsic mode frequencies are close enough to the SAW frequency. If they are not close, then no intrinsic mode may be excited because the efficiency of non-resonant excitation will be very poor. In that case, only the extrinsic mode will be excited and this seems to have happened in ref. \cite{drobitch}.

Because the  coupling of energy from the phonons in the SAW to the magnons in the spin waves becomes much more efficient if the excitation is resonant \cite{paper}, i.e., if the SAW frequency happens to match an intrinsic mode frequency, there will be {\it specific SAW frequencies} that will result in very large amplitude/power hybrid magneto-dynamical modes. They can radiate electromagnetic waves extremely efficiently. In other words, resonant excitation with SAW will be much more effective in generating electromagnetic waves than non-resonant excitation, and hence the radiation efficiency of the antenna will be much higher if the SAW excitation is resonant. 

The frequencies of the intrinsic modes depend on many parameters, including the direction of SAW propagation in the array, and are therefore difficult to calculate. However, they can be measured with such techniques as time-resolved magneto-optical Kerr effect microscopy, as was done in \cite{nanoscale}. In the context of antennas, if one were to measure the radiation efficiency of the antenna as a function of SAW frequency for a given direction of SAW propagation, {\it one would find exceptionally high efficiencies at some discrete frequencies}. These would be the intrinsic mode frequencies for that direction of SAW propagation, where very efficient energy transfer takes place from the SAW to the hybrid magnetodynamical modes and thence to electromagnetic waves. This three-way transfer is due to tripartite phonon-magnon-photon coupling where phonons from the SAW couple into magnons in the hybrid magneto-dynamical modes and the magnons then couple into photons in the electromagnetic wave. The coupling can be either resonant or non-resonant, with the former much stronger than the latter.

\begin{figure}[!h]
    \centering
    \includegraphics[width=0.99\linewidth]{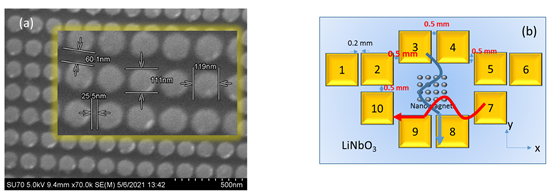}
    \caption{(a) Scanning electron micrograph of the nanomagnet array used in ref. \cite{raisa} showing the relevant dimensions and distances. (b) Ten electrodes are delineated around the nanomagnet array to launch a SAW in different directions. ``Orientation 1'' corresponds to the case when electrodes 3 and 4 are activated, while ``orientation 2'' corresponds to the case when electrodes 5 and 7 are activated. This ``antenna'' showed exceptionally high radiation efficiencies at three distinct SAW frequencies of 5, 14 and 15 GHz [depending on the direction of SAW propagation] most likely because those were the frequencies of the intrinsic modes of the array. Reproduced from \cite{raisa} with CC-BY license.}
    \label{fig:direction}
\end{figure}

Ref. \cite{raisa} observed this effect. It measured the radiation efficiencies at various SAW frequencies in an antenna made of the nanomagnet array shown in Fig. \ref{fig:direction} for two different directions of SAW propagation, and the results are tabulated in Table I. 

\begin{table}[!b]
\centering
    \caption{Radiation efficiencies at different SAW frequencies for SAW propagation in two different directions in the array shown in Fig. \ref{fig:direction}(a). The $A/\lambda_{EM}^2$ ratio is also plotted at each frequency. Reproduced from \cite{raisa} with CC-BY license.}
    \begin{tabular}{|c|c|c|c|}
    \hline
    {\bf SAW frequency (GHz)} & {\bf Orientation 1} & {\bf Orientation 2} & {\bf $A/\lambda_{EM}^2$} \\
    \hline
    5 & 0.356 & 0.116 & 2.9$\times10^{-6}$ \\
    10 & 0.0874 & & 1.16$\times10^{-6}$ \\
    13 & & 0.0725 & 1.97$\times10^{-5}$ \\
    14 && 0.544 & 2.28$\times10^{-5}$ \\
    15 & 0.149 & 0.121 & 2.61$\times10^{-5}$ \\
    18 & 0.00138 && 3.76$\times10^{-5}$ \\
    30 & & 0.047 & 1.04$\times10^{-4}$ \\
    35 && 0.0135 & 1.42$\times10^{-4}$ \\
        \hline
    \end{tabular}
\end{table}

\begin{figure}[!h]
        \centering
        \includegraphics[width=0.91\linewidth]{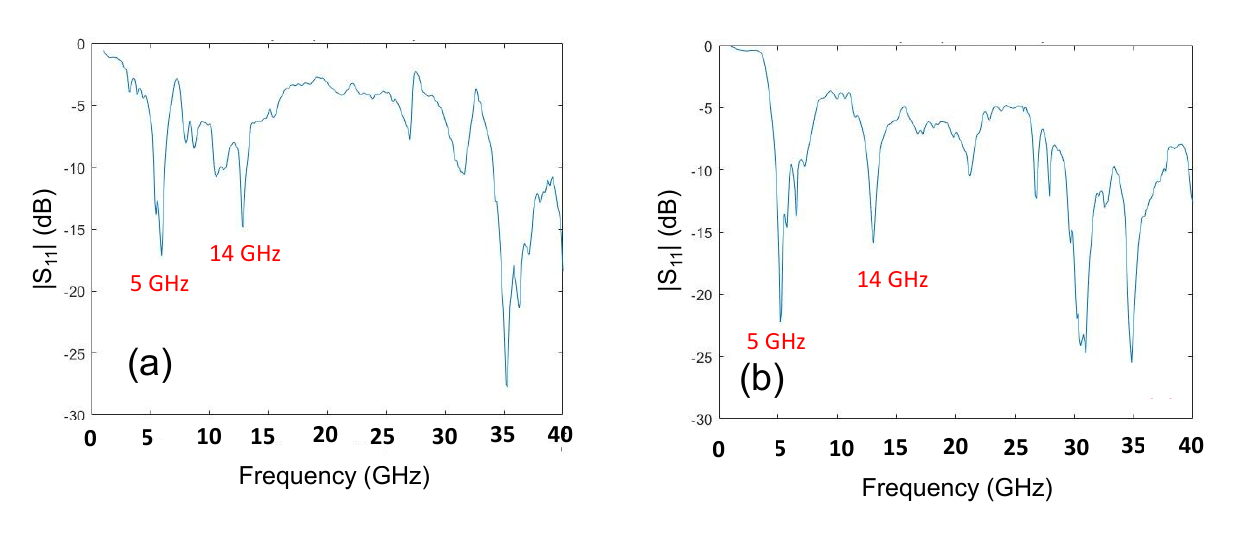}
        \caption{Spectra of the scattering parameter |S$_{11}$|, measured with a vector network analyzer, for two different directions of SAW propagation in the nanomagnet array of Fig. \ref{fig:direction}: (a) SAW propagating parallel to the major axes of the elliptical nanomagnets, and (b) SAW propagating parallel to the minor axes. Reproduced from \cite{raisa} with CC-BY license.}
        \label{fig:scattering}
    \end{figure}

The radiation efficiencies were found to be exceptionally high at three specific frequencies -- 5, 14 and 15 GHz -- and much lower at other frequencies. This is very likely due to the fact that these three frequencies are the intrinsic mode frequencies in the array [depending on the direction of SAW propagation] and when the SAW frequency matches them, the intrinsic modes are resonantly excited. These modes radiate relatively large power, making the radiation efficiency very high at these three frequencies. Resonant excitation (resonant phonon magnon-coupling) leading to strong electromagnetic radiation has been studied theoretically in \cite{paper}. 

Additional support for this conjecture can be gained from the spectra of the scattering parameter S$_{11}$ which is a measure of the fraction of the input microwave power fed to the SAW-launching electrodes that is reflected back into the source. This spectrum is shown in Fig. \ref{fig:scattering} and shows pronounced dips or troughs at 5, 14, 30 and 35 GHz [for two mutually perpendicular directions of SAW propagation] indicating that at these frequencies, the sample strongly absorbs SAW power and does not reflect it back so much into the source. This can happen only if efficient transfer of energy from the SAW to the hybrid magneto-dynamical spin wave modes is taking place at these frequencies. The absorbed power is used to generate intrinsic modes in the nanomagnets which then radiate electromagnetic waves very efficiently. 

This, of course, begs the question why we see high efficiencies at 5 and 14 GHz, but not at 30 and 35 GHz where the S$_{11}$ parameter has strong dips. One must note that there are two effects at play here: conversion of the SAW into hybrid magneto-dynamical spin waves via phonon-magnon coupling and then conversion of the spin waves into electromagnetic waves via magnon-photon coupling. While the former coupling may be efficient at 30 and 35 GHz as well (as the scattering parameter spectra suggest) the latter may not be. That could be why we see high radiation efficiencies at 5 and 14 GHz, but not at 30 and 35 GHz.

Note also that the radiation efficiencies at the intrinsic mode frequencies are five orders of magnitude higher than the $A/\lambda_{EM}^2$ limit that would have constrained traditional antennas.

\begin{figure}
    \centering
    \includegraphics[width=0.91\linewidth]{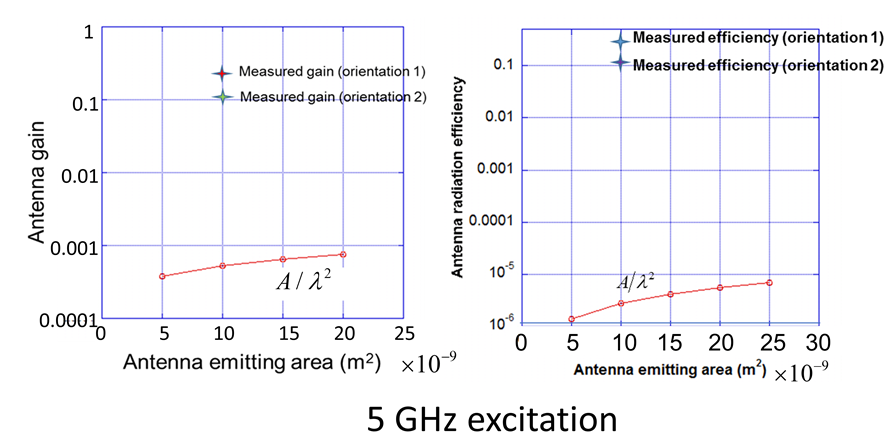}
    \caption{The $A/\lambda^2$ limit as a function of antenna emitting area and the measured gain and efficiency of the AMMC antenna at 5 GHz SAW excitation frequency. The two ``orientations'' correspond to two mutually perpendicular directions of SAW propagation. Reproduced from ref. \cite{raisa} with CC-BY license.}
    \label{fig:comparison}
\end{figure}

These antennas, based on {\it resonant} tripartite phonon-magnon-photon coupling, can produce exceptionally high radiation efficiencies at discrete frequencies that correspond to the intrinsic mode frequencies. Thus, they spawn a disruptive technology for embedded applications. A comparison between the $A/\lambda^2$ limit as a function of antenna emitting area and the measured radiation gain and efficiency for the AMMC antenna at 5 GHz SAW excitation frequency is shown in Fig. \ref{fig:comparison}. 
As a happy coincidence, the frequency of 5 GHz happens to be a popular WiFi frequency and that makes this case very attractive. Recently, these antennas have been modeled for their electromagnetic emission properties \cite{saibal1,saibal2}.

\subsection{More on phonon-magnon coupling in artificial multiferroic magnonic crystals}

We call the two-dimensional periodic array of nanomagnets on a piezoelectric substrate an ``artificial multiferroic magnonic crystal'' (AMMC) for two reasons. First, the periodicity of the array imposes periodic boundary conditions on any traveling spin waves generated in them and therefore mimics a two-dimensional ``crystal''. The term ``magnonic crystal'' has been used by many authors in this context, and its application in data processing has been widely explored \cite{chumak}. Here, we have a different application, namely in wireless communication. Second, the coupling between the magnetostrictive phase and the piezoelectric phase mimics a multiferroic. Hence the name ``artificial multiferroic magnonic crystal'' (AMMC).

The reason why surface acoustic waves excite hybrid magneto-dynamical spin wave modes in an AMMC is of course because of phonon-magnon coupling which would require a rigorous quantum-mechanical treatment. However, to understand this qualitatively within a classical framework, one can think of the SAW (or time varying strain) as producing an effective time-varying magnetic field within a magnetostrictive nanomagnet owing to magneto-elastic coupling \cite{nanoscale}.  This magnetic field has the same frequency as the SAW, is directed along the direction of SAW propagation, and its amplitude is related to the following parameters: (1) SAW power per unit area in the plane perpendicular to the direction of SAW propagation, (2) the acoustic impedance of the medium, (3) the properties of the substrate, and (4) the saturation magnetostriction of the nanomagnets \cite{nanoscale,raisa}. Typically, the amplitude is a few tens of Oersted \cite{raisa}. This  field excites the hybrid magneto-dynamical spin wave modes within the nanomagnets.

\begin{figure}[!h]
\vspace{-0.4cm}
    \centering
    \includegraphics[width=0.99\linewidth]{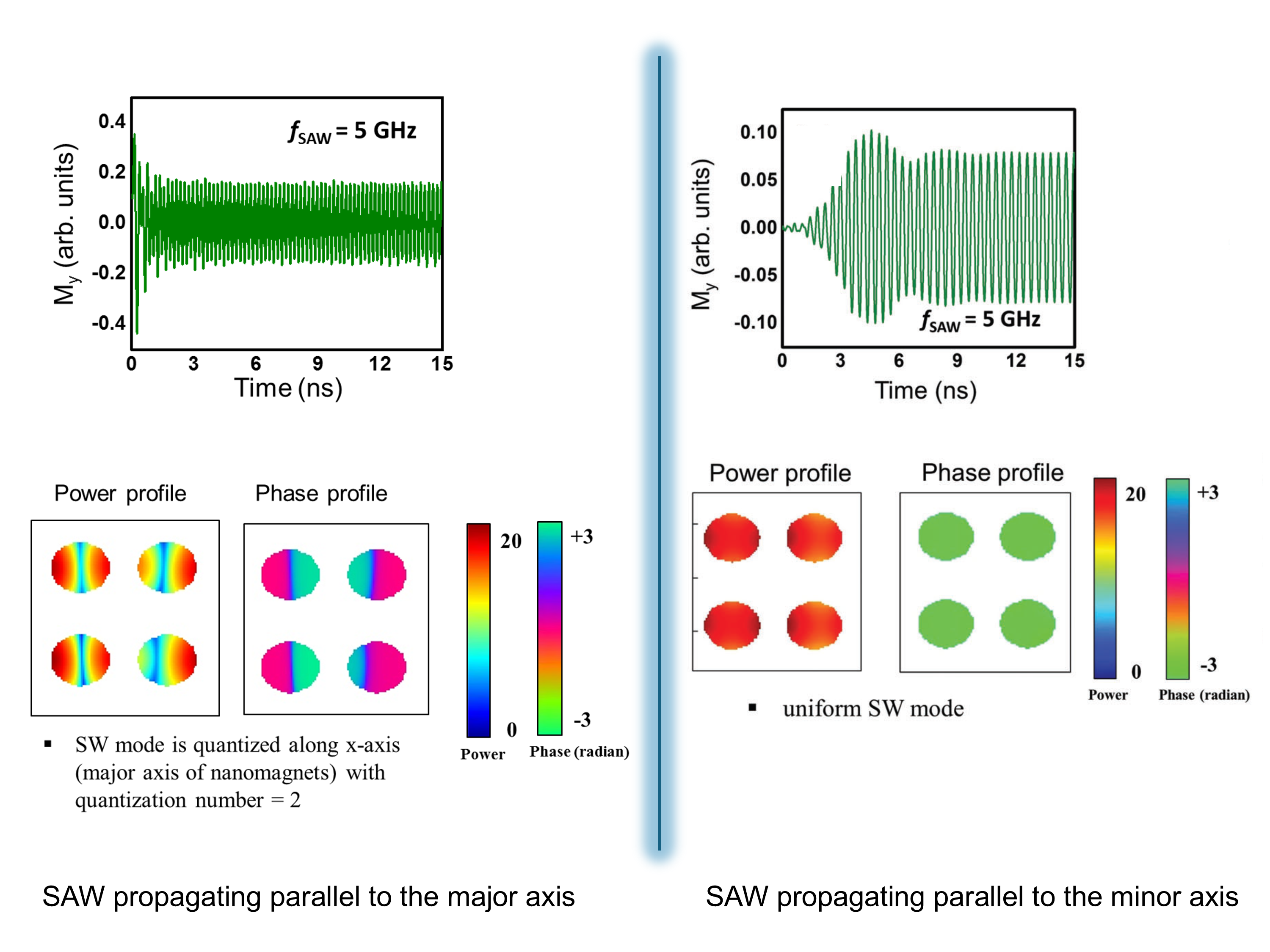}
    \caption{Spin wave pattern (oscillation of the y-component of magnetization where the y-direction is indicated in Fig. \ref{fig:direction}) as well as power and phase profiles calculated at 5 GHz SAW frequency.}
    \label{fig:directional}
\end{figure}

Ref. \cite{raisa} calculated the magnetization oscillation waveforms and the power/phase profiles of the hybrid magneto-dynamical modes generated in the AMMC shown in Fig. \ref{fig:direction} at various SAW frequencies using the Landau-Lifshitz-Gilbert equations. The power and phase profiles were expectedly found to depend on the direction of SAW propagation since that determines the direction of the effective time-varying magnetic field. Furthermore, the two-dimensional array does not have rotational symmetry in its plane, which also contributes to the directional dependence. The magnetization oscillation waveforms as well as the power and phase profiles of the hybrid magneto-dynamical modes produced in the AMMC in Fig. \ref{fig:direction} are shown in Fig. \ref{fig:directional} for the case when the launched SAW frequency is 5 GHz (one of the intrinsic mode frequencies). Clearly, all quantities depend on the direction of SAW propagation. 

Ref. \cite{raisa} also reported Fourier transforms of the magnetization oscillation waveforms to extract the frequency components of the hybrid magneto-dynamical modes for two different directions of SAW propagation (parallel to the major and the minor axes of the elliptical nanomagnets). The dominant frequency is of course almost always the SAW excitation frequency, but there are also other frequencies that are not necessarily integral multiples of the excitation frequency. These are probably the frequencies of the intrinsic modes that are off-resonantly excited by the SAW. The spectrum of the oscillations also depends on the direction of SAW propagation for any SAW frequency because the intrinsic modes depend on that. Ref. \cite{raisa} then went on to investigate if this directional dependence of the magnetization oscillation spectra made the spectrum of the emitted electromagnetic radiation also depend on the direction of SAW propagation at any SAW frequency. It did. The dominant frequency of the emitted radiation is almost always the launched SAW frequency as expected, but there are sidebands [again due to intrinsic modes that are off-resonantly excited by the SAW] whose frequencies and amplitudes vary if the direction of SAW propagation is varied. Thus, one can change the spectra of the emitted electromagnetic radiation by changing the direction of SAW propagation in the AMMC. This is an unusual and intriguing property.

We conclude this section by pointing out that phonon-magnon coupling in AMMC-type structures has been an active field of research \cite{NPG,babu,berk,JAP,schmidt,vidal}. However, its application has been lacking. The antenna application is one of the first, if not the first. Recently, there has also been an application of this phenomenon in mode-selective amplification of spin waves \cite{amplifier}. More applications may be around the corner.

\section{Beam steering with directed surface acoustic wave in the extreme sub-wavelength AMMC antenna}
\label{sec:beam}

The dependence of the magnetization oscillation waveforms on the direction of SAW propagation in the  AMMC antenna [shown in Fig. \ref{fig:directional}] can be exploited to achieve an important technological feat -- ``beam steering'' with a {\it single} antenna element much smaller than the wavelength. Beam steering, which is at the heart of active electronic scanning arrays (AESA), usually requires a {\it phased array} comprising multiple antenna elements, each much larger than the wavelength. This large size makes a beam steerer unwieldy and unimplantable. The stumbling block was the fact that classically if an antenna is much smaller than the wavelength, then it is effectively a ``point source'' which should radiate omnidirectionally and isotropically, i.e., the radiated power will be the same in all directions. Such an antenna would not even  form a beam, let alone allow for the beam to be steered. This is why a phased array is needed for beam steering and consequently miniaturization has been elusive.

\begin{figure*}[!h]
\centering
\includegraphics[width=0.99\textwidth]{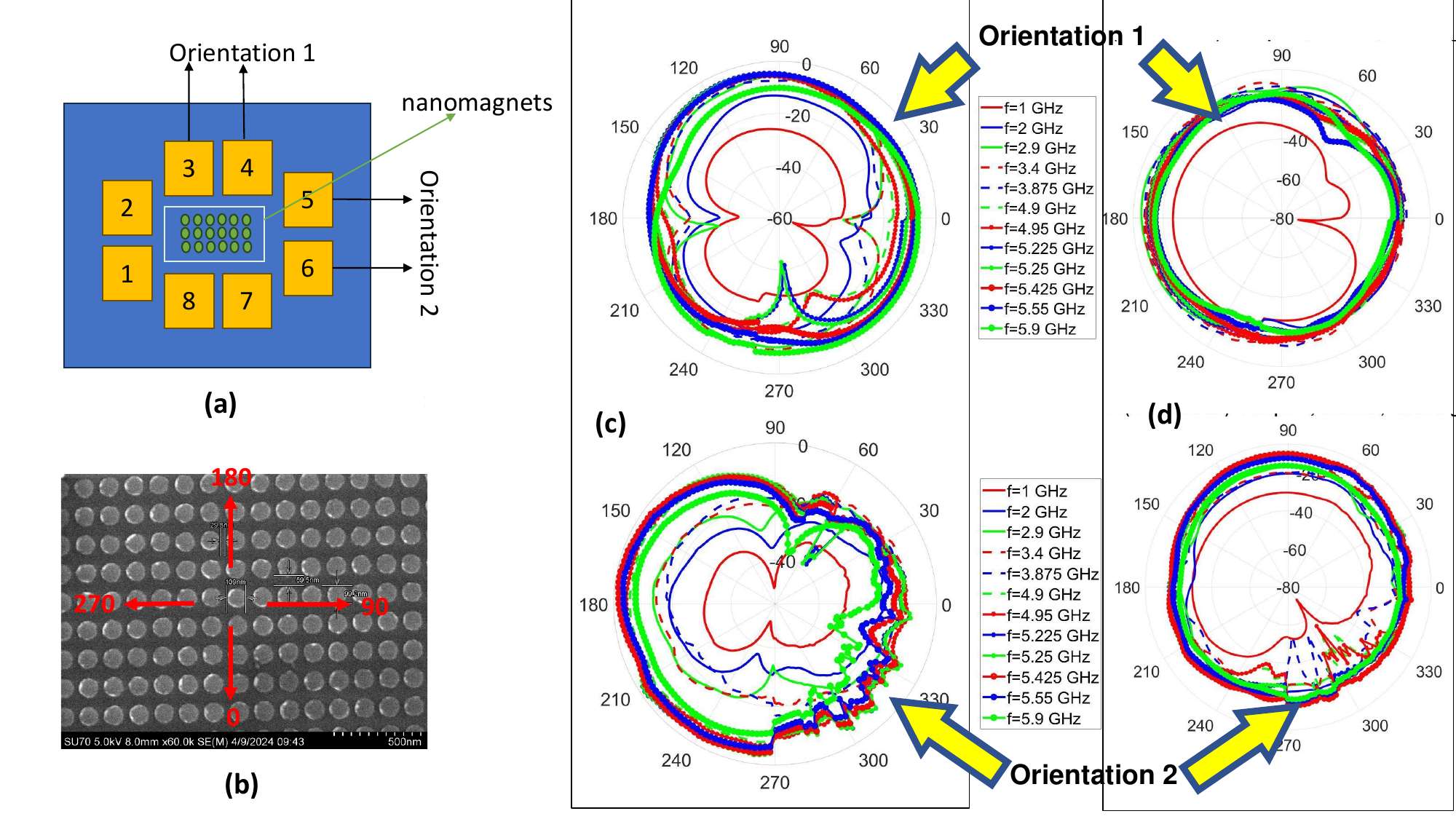}
\caption{Radiation patterns (gain in dbi) in the {\it plane of the nanomagnets}  at different SAW excitation frequencies when the microwave source to launch the SAW is connected between two different electrode pairs in order to make the SAW propagate along two different directions. This is the same sample as that shown in Fig. \ref{fig:direction}. (a) ``orientation 1'' is defined as the case when the microwave source is connected between electrodes 3 and 4 to launch a SAW propagating parallel to the major axes of the elliptical nanomagnets, while ``orientation 2'' refers to the case when the microwave source is connected between electrodes 5 and 6 to launch a SAW propagating parallel to the minor axes of the elliptical nanomagnets (b) Definition of the ``directions'' for the radiation pattern; 0$^{\circ}$ is along the minor axes and 90$^{\circ}$ is along the major axes of the nanomagnets. (c) Radiation patterns for horizontal polarization for both ``orientations''. (d) Radiation patterns for vertical polarization for both ``orientations''. Note that the radiation pattern is ``orientation''-dependent, meaning that it depends on the direction of SAW propagation. Reproduced from \cite{steering} with permission of the Institute of Electrical and Electronics Engineers.}
\label{fig:pattern1}
\end{figure*}

Fortunately, the AMMC antenna does {\it not} radiate omnidirectionally or isotropically, despite being effectively a point source, because of its ``internal anisotropy'' accruing from the dependence of the magnetization oscillation waveforms on the direction of SAW propagation as shown in Fig. \ref{fig:directional}. Not only does the spectrum of the emitted radiation in such an antenna depend on the direction of SAW propagation, as observed in Ref. \cite{raisa}, but it turns out that the {\it radiation pattern} in any plane (i.e., the directional dependence of the radiation intensity in any plane) also depends on the direction of SAW propagation \cite{steering}.

Fig. \ref{fig:pattern1} shows the radiation pattern of the AMMC antenna in Fig. \ref{fig:direction} (measured in the plane of the nanomagnets) for two different directions of SAW propagation [parallel to the major and minor axes of the elliptical nanomagnets] labeled as ``orientation 1'' and ``orientation 2''. These far-field patterns were measured in an anechoic chamber with a polarization-sensitive horn antenna to yield the patterns for two orthogonal polarizations termed ``horizontal'' and ``vertical''. The radiation patterns in the two planes that are transverse to the plane of the nanomagnets can be found in ref. \cite{steering}.

\begin{figure}[!b]
    \centering
    \includegraphics[width=0.91\linewidth]{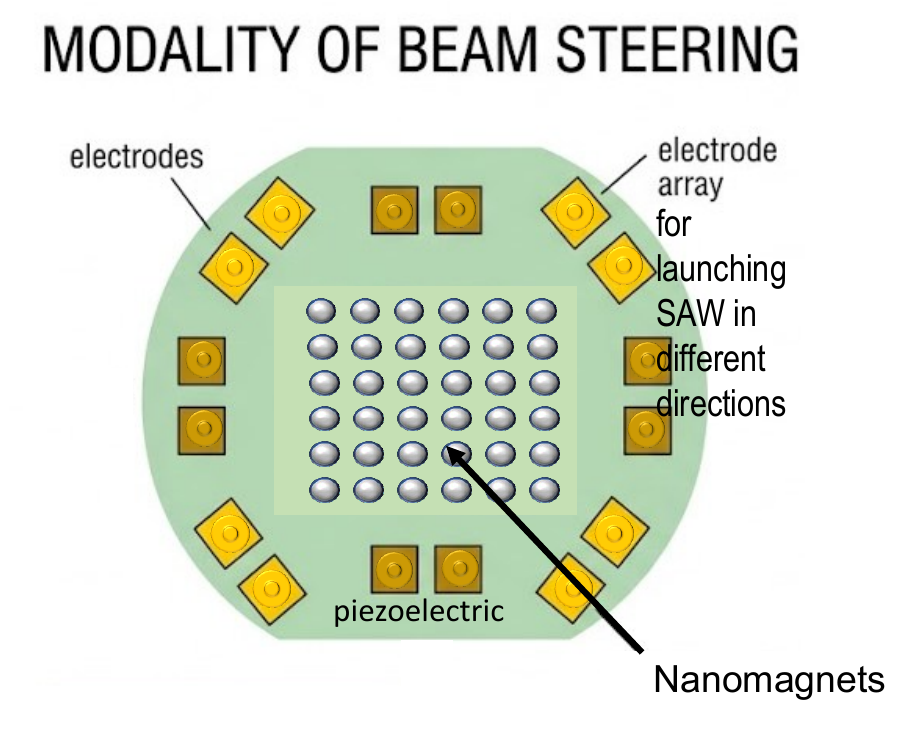}
    \caption{A scheme for continuously steering the radiated beam or rotating the radiation pattern  in any plane. Adjacent electrode pairs are sequentially activated with a multiphase clock to scan the principal lobe of the radiation pattern over 360$^{\circ}$ in any plane and attain AESA functionality.}
    \label{fig:electrodes}
\end{figure}

Clearly, the radiation patterns depend on the direction of SAW propagation for either polarization and hence we can change them by changing the direction of SAW propagation. This effectively enables beam steering. Interestingly, at 1 GHz, if we rotate the direction of SAW propagation by 90$^{\circ}$, the radiation pattern also rotates clockwise by about 90$^{\circ}$ for the horizontal polarization and by about 300$^{\circ}$ counter-clockwise for the vertical polarization. At higher frequencies, the radiation pattern becomes more isotropic and hence this feature gets subdued when the frequency increases.

The exciting result here is that one is able to alter the radiation pattern {\it using a single antenna element much smaller than the wavelength}, without employing a phased array which typically has multiple antenna elements each much larger than the wavelength. This enables miniaturization and therefore embedded applications. 

In principle, one can continuously change the direction of SAW propagation by sequentially activating different adjacent (SAW-launching) electrode  pairs surrounding the nanomagnet array with a multiphase clock and thus be able to steer the main lobe of the radiation pattern (if there is one) continuously over an entire plane to attain AESA functionality. This is schematically shown in Fig. \ref{fig:electrodes}. The only drawback is that the radiation pattern does not remain intact and gets somewhat distorted during the rotation, but this may not be completely unacceptable in some scenarios.

\section{Polarization switching ultra-sub-wavelength stealthy antenna for polarization division multiplexing and secure or  covert communication}

The polarization sensitivity of the radiation pattern gave way to another idea. 
In quantum communication, information is encoded in the {\it polarization} of a photon for quantum key distribution and other tasks that demand security. In classical communication, encoding information in the polarization of an electromagnetic wave offers some advantages as well, such as polarization encoded secret sharing \cite{oe,li2} and  multiple data streams sent on the same frequency channel using different polarization states to save bandwidth. The latter is known as {\it polarization division multiplexing}.

For wireless {\it digital} data transmission via two orthogonal polarizations of a transmitted electromagnetic wave that will encode bits 0 and 1, one must be able to switch the polarization of the emitted  wave from approximately horizontal (encoding the bit 0) to approximately vertical (encoding the bit 1), or vice versa. Absolute polarization purity is not a concern for such applications. As long as the two polarizations are distinguishable from each other (i.e., they are approximately orthogonal), it will suffice. 

Microwave polarization switches, i.e., antennas that transmit microwaves whose polarizations are switched between the two orthogonal states at will, are rare since they are usually much larger than the microwave wavelength (1-10 cm) and hence not integrable on a chip, although there are some near-exceptions \cite{staacke,jin}. Consequently, most  polarization switches work at optical frequencies where they are still larger than the optical wavelength ($\sim$1 $\mu$m) but small enough to be integrated on a chip. 

The AMMC antenna just described is extremely versatile; it can break the Harrington limit overwhelmingly, it can steer a beam with a single antenna element cmuch smaller than the wavelength, and it can also work as a microwave polarization switch antenna that  is orders of magnitude {\it smaller} than the operating wavelength. This allows integration on a chip. That can have multiple applications, such as in quantum computing \cite{staacke,brown}, remote sensing \cite{tyo} and, most importantly, aggressively miniaturized  ultra-compact polarization division multiplexers and (polarization-encoded) digital data transmitters in the microwave range. Additionally, it turns out that they may have other important applications, such as in secure communication in embedded applications (e.g., medically implanted devices, tiny stealthy drones for defense and crime-fighting), stealthy antennas, and physical unclonable functions (PUF) for antenna authentication and trust \cite{covert}. 

To elucidate polarization encoding in this antenna, we first define a so-called ``polarization angle''. Let us say that at a given frequency, the horizontal polarization component of the emitted wave in any direction  is $h$ and the vertical component is $v$. We define a polarization angle $\theta$ in that direction as 
\begin{equation}
    \theta = tan^{-1} \left ( \frac{v}{h} \right ).
\end{equation}
Here, $\theta$ = 0 corresponds to purely horizontal polarization and $\theta$ = $\pi/2$ radians corresponds to purely vertical polarization.

\begin{figure}[!hbt]
\centering
\includegraphics[height=5.5in,angle=270]{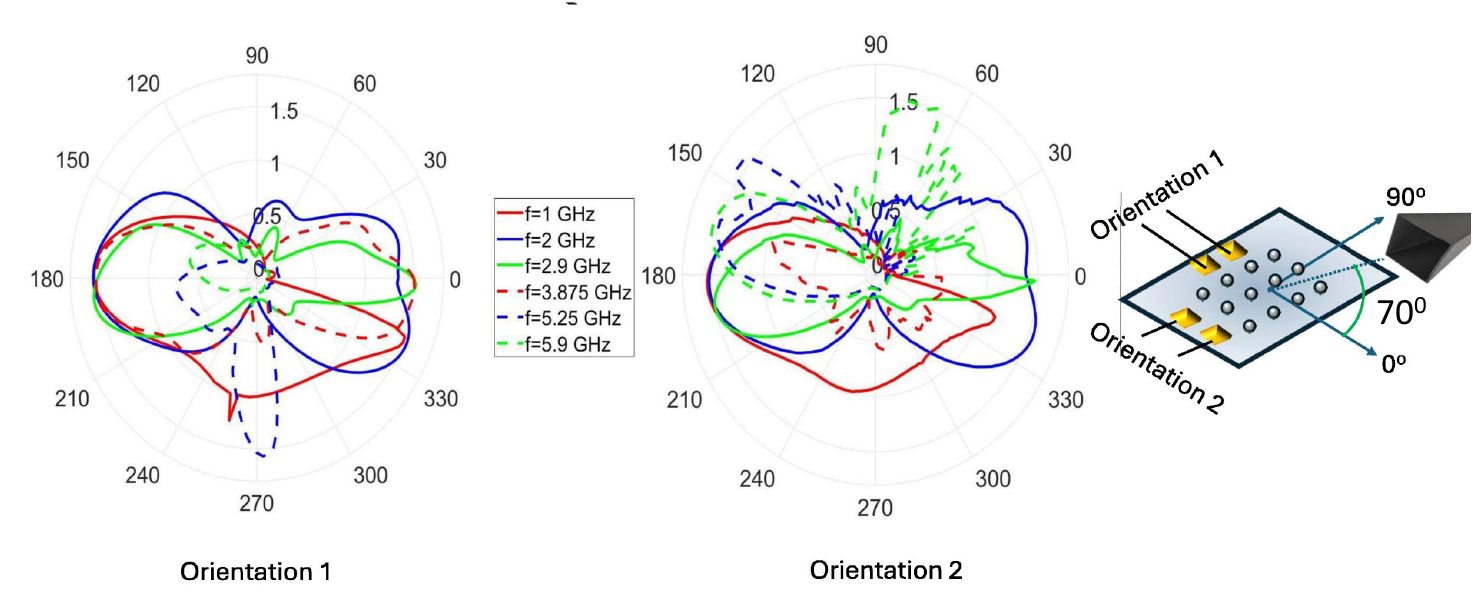}
\caption{Radial plot of the polarization angle $\theta$ (in radians) in the plane of the nanomagnets at different frequencies for two mutually perpendicular directions of SAW propagation labeled as ``orientation 1'' and ``orientation 2''. Again, this is the same sample as that shown in Fig. \ref{fig:direction}. The far right panel shows that only if we place the receiving antenna at 70$^{\circ}$ angle in the plane of the nanomagnets and transmit at 5.9 GHz frequency, then we will receive and detect two mutually orthogonal polarizations if we switch the direction of SAW propagation from ``orientation 1'' to ``orientation 2'' by switching the electrode pairs used to launch the SAW. This enables polarization-encoded digital transmission {\it only at this frequency and only in this direction}. Reproduced from \cite{covert} with CC-BY license.}
\label{fig:Fig4}
\end{figure}

In Fig. \ref{fig:Fig4}, we show the radial plot of $\theta$ at different frequencies in radians in the plane of the nanomagnets. Note that at 5.9 GHz frequency and in the 70$^{\circ}$ direction, $\theta$ is nearly $\pi/2$ radians for orientation 2 and nearly 0 radians for orientation 1. Thus, by switching the direction of SAW propagation from orientation 1 to orientation 2, we can change the polarization of the beam radiated in the 70$^{\circ}$ direction at 5.9 GHz (in the nanomagnets' plane) from nearly {\it horizontal} to nearly {\it vertical} and thus transmit bits 0 and 1, but {\it only in that direction and only at that frequency}.  We can place the receiver at 70$^{\circ}$ angle in the plane of the nanomagnets as shown in the right panel of Fig. \ref{fig:Fig4}, transmit at 5.9 GHz, and receive either nearly horizontal polarization or nearly vertical polarization and no ambiguous polarization at the receiver. This enables polarization-encoded digital communication. Note that the positions of the transmitting device and the receiver do not have to be fixed in space, but their relative alignment has to be fixed in this scheme. This enables {\it point-to-point} communication via polarization.

Consider yet another frequency of 5.25 GHz and the 275$^{\circ}$ direction in the nanomagnets' plane. For orientation 1, $\theta$ is very close to $\pi$/2, meaning that the emitted beam is vertically polarized when the SAW is launched in the direction corresponding to orientation 1, whereas if the SAW is launched in the direction corresponding to  orientation 2, $\theta$ is close to 0, meaning that the emitted beam is horizontally polarized. This is the opposite of the previous case because here orientation 1 corresponds to transmitting bit 1 and orientation 2 corresponds to bit 0. Thus, there are multiple directions and frequencies at which point-to-point communication is enabled via polarization encoding. This feature also enables {\it mutiple-input-multiple-output} (MIMO) antenna functionality. Multiple bit streams, each with its own carrier frequency, can be transmitted in multiple different directions simultaneously to improve reliability, countering signal fading, etc.

The specific directions and the specific frequencies for communication (70$^{\circ}$ and 5.9 GHz, or 275$^{\circ}$ and 5.25 GHz in this case) will depend on the antenna parameters such as size, shape, spacing, material composition, etc. of the nanomagnets and can be changed by changing these parameters.  

\subsection{Covert communication}

The fact that the point-to-point communication via polarization encoding works only for {\it a specific alignment of
the transmitter and receiver} makes {\it covert} communication \cite{jiang,nan} possible. The authorized sender has access to the polarization
radial plot and can inform the authorized receiver where to place the receiving device (i.e., provide the polar and the azimuthal angle). An unauthorized receiver or eavesdropper will not know where to place the receiver and will place it at a
wrong location with very high probability, thereby not receiving nearly pure binary polarization states. Hence the eavesdropper will be unable to decipher the transmitted message. This is essentially {\it steganography} \cite{fridrich,grabsky}. It also eliminates the need for cryptography which is 
always vulnerable to sophisticated attacks.

\subsection{Physical unclonable function}

The scheme described here is also {\it unclonable} since the specific alignments and frequencies that work will vary slightly from sample to sample because
of unavoidable manufacturing variations involved in the fabrication of the AMMC antenna. Thus, the correct alignment direction and frequency are a ``biometric'' of any chosen antenna. The
antenna acts as a physical unclonable function (PUF) that cannot be reproduced or predicted \cite{devadas}, thereby providing
strong security.

\subsection{Stealthy antenna}

Finally, the above feature introduces an element of {\it stealth} in the antenna operation. Correct bit streams are transmitted only the right direction at the right frequency and an eavesdropper whose receiver is not placed in the right location and tuned to the right frequency will receive garbled bit streams. For example, at 5.9 GHz, $\theta$ = 0 for both directions of SAW propagation (both orientations 1 and 2) in the 340$^{\circ}$ direction. Hence, a receiver placed in that 340$^{\circ}$ direction will receive a constant stream of 0's if the horizontal polarization encodes the bit 0. This bit stream contains no message. Similarly, in the 165$^{\circ}$ direction, $\theta$ = $\pi/2$ radians for orientation 2 and 0.61 radians for orientation 1 at 5.9 GHz. Thus, the beam is 100\% vertically polarized for the choice of orientation 2 while being 67\% horizontally polarized and 33\% vertically polarized for the choice of orientation 1 if the chosen frequency is again 5.9 GHz. The latter impure polarization states do not encode binary bits unambiguously and hence carry no meaningful message.  Thus, the message is concealed from anyone without precise knowledge of the correct orientation and frequency. This enables stealth.

\section{Quantum-enabled ultra-sub-wavelength nanomagnetic antenna -- with beam steering capability -- based on spin injection into nanomagnets from a topological insulator}

The extreme-sub-wavelength AMMC antennas require a multiferroic system -- magnetostrictive nanomagnets elastically coupled to an underlying piezoelectric substrate. There are, however, other types of similar miniaturized antennas of ultra-sub-wavelength dimensions that radiate with radiation efficiencies several times higher than the Harrington limit and may even allow beam steering with a single element much smaller than the electromagnetic wavelength. They are different from the AMMC antenna in that they do not require either magnetostriction or piezoelectricity, and do not operate on the basis of phonon-magnon-photon coupling. Instead, they employ other unconventional principles which are also of spintronic origin and excite magnetization oscillation in nanomagnets (again, spin waves radiating electromagnetic waves). What is attractive about them is that they could be fabricated on a silicon substrate and hence will be fully compatible with silicon technology. We will describe two of them in this section and the next.

\subsection{An antenna implemented with nanomagnets deposited on the surface of a three-dimensional topological insulator}

A three-dimensional topological insulator (3D-TI) is a special class of quantum materials which is ideally insulating in the bulk but conducting on the surface. The surface electrons (charge carriers) are ``spin-momentum locked'' in the sense that the spin is always polarized perpendicular to the direction of the electron's momentum. If the momentum's direction changes, then the spin polarization changes with it \cite{moore,li,kang}.

\begin{figure}[!h]
    \centering
    \includegraphics[width=0.99\linewidth]{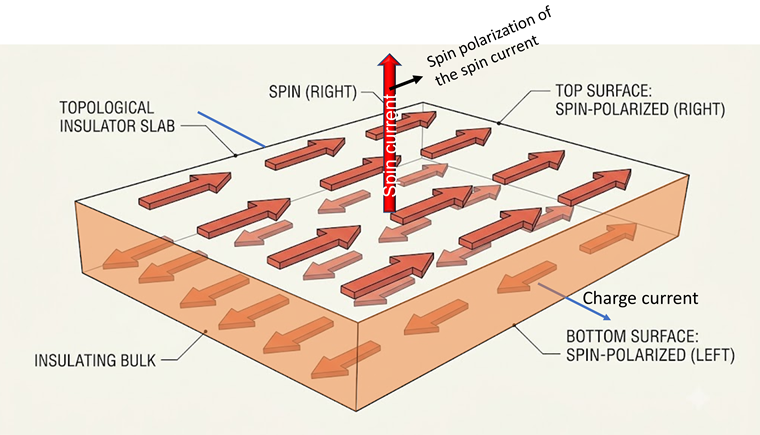}
    \caption{A 3D-TI into which a charge current is injected. The top and bottom surfaces become spin-polarized in opposite directions resulting in the flow of a spin current perpendicular to the surfaces. If a magnet is placed on the top surface, the spin current will inject spins polarized in the direction of the top surface's spin polarization into the magnet. Note that the spin current, charge current and the injected spin are mutually perpendicular. }
    \label{fig:TI}
\end{figure}

If a charge current is injected into a slab of a 3D-TI, then there is a net electron momentum in the direction of the current. The top and bottom surfaces become spin-polarized in opposite directions that are transverse to the direction of the charge current because of spin-momentum locking. This causes a spin current to flow perpendicular to the surfaces as shown in Fig. \ref{fig:TI}. That spin current can inject spins into a nanomagnet placed on the top surface \cite{mellnik,wang}. The spin {\it polarization} of the injected spin current is the same as that of the top surface and hence it is perpendicular to both the spin current and the charge current. 

Because of spin-momentum locking, if the direction of the charge current flow is reversed, then the spin polarization of the spin current should be reversed as well. Therefore, if an {\it alternating} charge current is injected into a 3D-TI, it will inject a spin current of {\it alternating} spin polarization into a nanomagnet placed on the surface. This will exert an alternating spin-orbit torque (SOT) on the nanomagnet,  causing spin waves to be generated within it \cite{morrison}. These are not the hybrid magneto-dynamical modes since they are not born of phonon-magnon coupling but they can also transfer energy to electromagnetic waves via magnon-photon coupling and radiate these waves into the surrounding medium \cite{kuo} with the frequency of the charge current. This implements an ``antenna'' \cite{TI-antenna}. 

The efficiency of energy transfer from the spin waves to the electromagnetic waves (and hence the radiation efficiency of the antenna) depends on the strength of magnon-photon coupling, which can be quite strong in some circumstances \cite{xiao,rao}, particularly in the presence of interface spin-orbit torque (SOT) \cite{salikhov}.

Besides implementing an unconventional antenna, the rich physics of topological insulators, particularly {\it spin-momentum locking},  enables beam-steering as well. Because of the fixed angular relationship between the axis of the alternating spin polarization of the spin current and the direction of the charge current density (due to spin-momentum locking) [see Fig. \ref{fig:TI}], we can rotate the axis of the oscillating spins through any angle by rotating the direction of the alternating current flow. This will then rotate the spin wave pattern in space and therefore alter the electromagnetic {\it radiation pattern}, thereby effectively implementing ``beam steering''. A phenomenological proof of this effect can be found in the  supplementary section of Ref. \cite{TI-antenna}.

\begin{figure}[!h]
\centering
\includegraphics[width=0.99\textwidth]{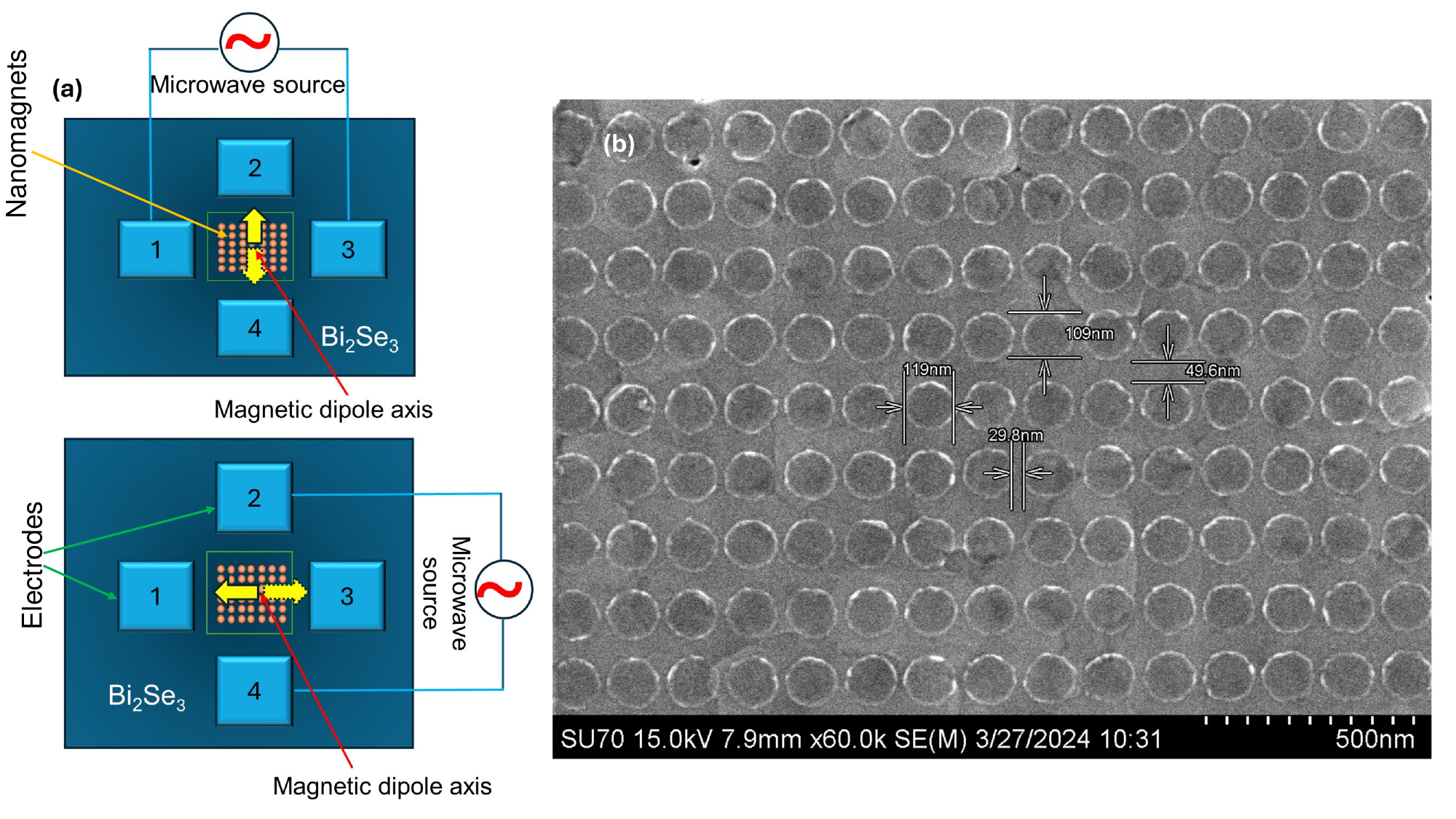}
\caption{(a) Schematic of the nanomagnet/TI based antenna and two different ways of passing an alternating charge current between antipodal electrode pairs leading to two different directions of charge current flow and hence two different axes of spin polarization in the spin current injected into the nanomagnets. The axes of the oscillating spins are shown for the two cases. The top configuration is referred to as ``orientation 1'', and the bottom as ``orientation 2''. (b) Scanning electron micrograph of the fabricated nanomagnet array. There are 15.36 million nanomagnets in the array, making the total antenna area about 0.003 cm$^2$. Reproduced from \cite{TI-antenna} with permission of the American Institute of Physics.}
\label{fig:TI-antenna}
\end{figure}

The experiment reported in Ref. \cite{TI-antenna} used the sample configurations shown in Fig. \ref{fig:TI-antenna}(a) to demonstrate the antenna behavior. The sample consisted of a two-dimensional periodic array of $\sim$15 million slightly elliptical Co nanomagnets deposited on a 20-nm thick Bi$_2$Se$_3$ film (a 3D-TI).  The major axis of a nanomagnet was $\sim$120 nm, the minor axis was $\sim$110 nm and the thickness was 6 nm with a 5 nm Ti adhesion layer underneath each nanomagnet. 
The edge-to-edge separation between nearest neighbors  was $\sim$30 nm along the major axes of the nanomagnets and $\sim$50 nm along the minor axis. A scanning electron micrograph of the nanomagnet array with all relevant dimensions is shown in Fig. \ref{fig:TI-antenna}(b).

\subsection{Measurement results}

To test the nanomagnet/3D-TI samples, an alternating charge current of 1- 10 GHz frequency  was injected between either electrode pair 1 and 3, or between electrode pair 2 and 4 in Fig. \ref{fig:TI-antenna}(a), to inject spin currents of alternating spin polarization  into the nanomagnets. That generated alternating spin-orbit torque on the nanomagnets, which excited spin waves in them that radiated electromagnetic waves in the surrounding medium.

The spin oscillation axis in the nanomagnets for either electrode pair activation is shown in Fig. \ref{fig:TI-antenna}(a).   Changing this axis by changing the direction of the charge current flow changed the spin wave pattern within the nanomagnets and therefore the electromagnetic radiation pattern in any plane. This resulted in beam steering.  This is a new modality of beam steering made possible by spin-momentum locking of surface electrons in a 3D-TI and is therefore of quantum-mechanical origin since spin-momentum locking is a quantum-mechanical attribute. It leads to an ultra-miniaturized beam scanner consisting of a single antenna much smaller than the wavelength. This beam scanner is fundamentally different from the one described in section \ref{sec:beam} (the AMMC beam scanner) but serves the same purpose.

\subsection{Radiation patterns} 

Ref. \cite{TI-antenna} reported the radiation patterns of the nanomagnet/TI sample in three different planes -- the plane of the nanomagnets and the two transverse planes.  The patterns were measured at frequencies of 3.4, 5 and 10 GHz (free space wavelengths of 3, 6 and 9 cm) in an AMS-8100 anechoic chamber using a polarization-sensitive horn antenna placed far enough away from the sample to measure the far-field pattern at all frequencies. The input power (associated with charge current injection) was 1 mW.

Once again, two sets of samples were fabricated that were nominally identical, except one contained nanomagnets and the other did not. Following the usual convention, the first was termed ``real sample'' and the second was termed ``control sample''. This was done to confirm that the nanomagnets were radiating. If they were, then there would be a measurable difference between the radiation patterns of the two samples.
\begin{figure}[!hbt]
\centering
\includegraphics[width=0.91\textwidth]{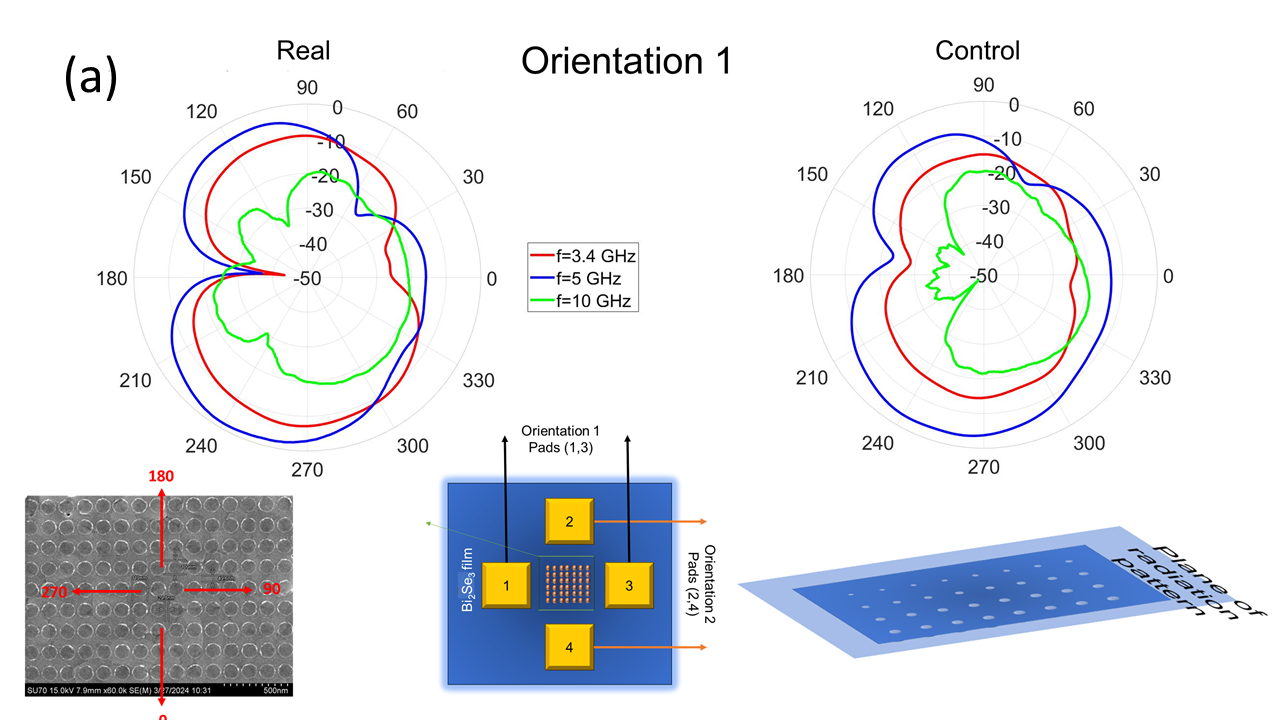}
\includegraphics[width=0.91\textwidth]{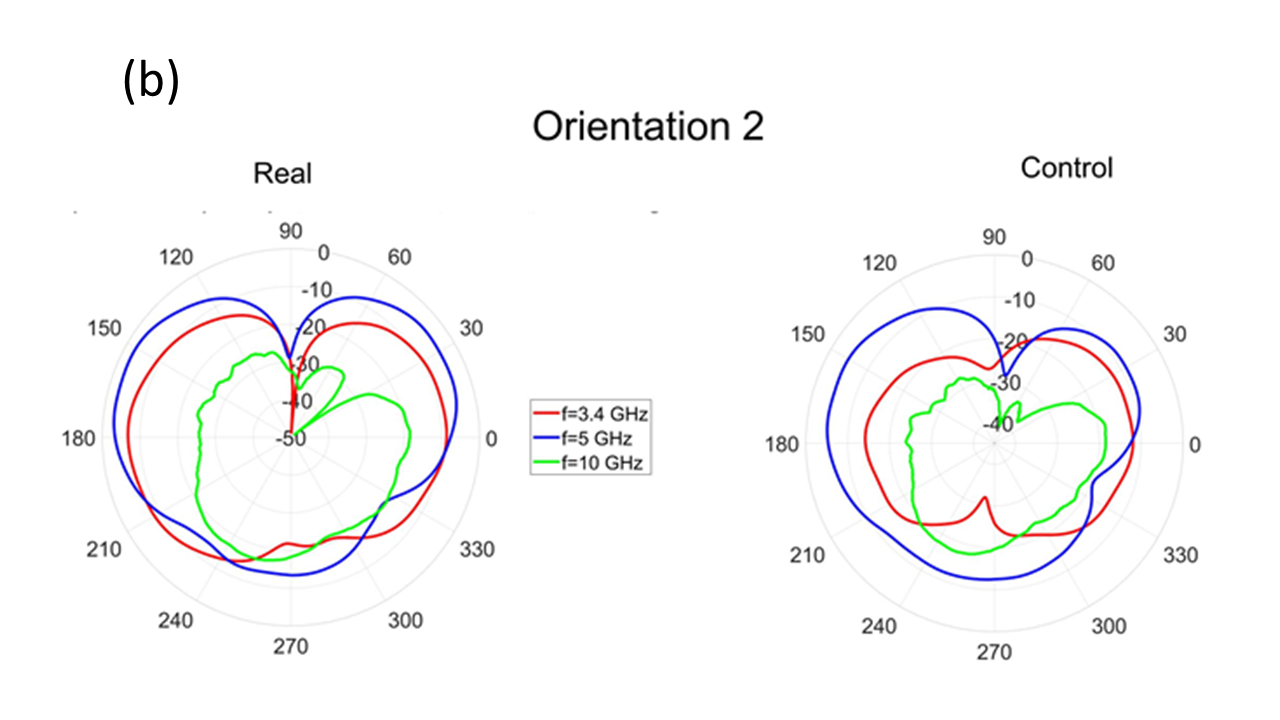}
\caption{The radiation patterns for the horizontal polarization of the electromagnetic wave emitted by the nanomagnet/TI sample in ref. \cite{TI-antenna} at three different frequencies (3.4, 5 and 10 GHz) in the plane of the nanomagnets. The detector was placed at a distance of 284.5 cm  from the sample. The patterns are shown for both the real sample and the control sample for two different directions of charge current flow -- (a) orientation 1 and (b) orientation 2, which are depicted in the inset. The patterns are plotted in dbi.  Reproduced from \cite{TI-antenna} with permission of the American Institute of Physics.}
\label{fig:control}
\end{figure}

Fig. \ref{fig:control} shows the radiation patterns in the plane of the nanomagnets reported in ref. \cite{TI-antenna} for both the real and the control sample, for horizontal polarization of the emitted radiation, and for both directions of alternating charge current injection (orientations 1 and 2). Strangely, in some directions in the plane of the nanomagnets (e.g., the 45$^{\circ}$ direction), the received power from the control sample at the location of the detector turned out to be {\it more} than that from the real sample, which might seem to suggest that the nanomagnets are not radiating power, but instead absorbing power! This apparent anomaly can be explained by the phenomenon of {\it destructive interference}. The real sample contains nanomagnets and peripherals (contacts pads, connectors, etc.) while the control sample contains only the peripherals. The waves transmitted by the peripherals and those transmitted by the nanomagnets in the real sample interfere at any point in space. In some directions, the interference at the location of the receiving horn antenna is {\it destructive}, which makes the power received from the real sample less than that from the control sample. This is the cause of the apparent anomaly.

However, in some other directions, the power received from the real sample {\it far exceeded} that from the control sample, which is a direct confirmation that the nanomagnets are radiating. For example, at 10 GHz and orientation 1 [see Fig. \ref{fig:control}], the measured gain in the 225$^{\circ}$ direction [horizontal polarization] was about -50 dbi for the control sample and -20 dbi for the real sample. This difference of 30 dbi (a factor of 1000$\times$) is large enough to confirm that the nanomagnets were definitely radiating, at least in the 225$^{\circ}$ direction at 10 GHz, and perhaps in other directions as well.

What these features also tell us is that the radiation from the nanomagnets is extremely {\it directional} or anisotropic because the difference between the radiation from the real sample and the control sample is strongly direction-dependent. This strong anisotropy is unexpected since the  entire nanomagnet array is orders of magnitude smaller than the free-space electromagnetic wavelength at all measurement frequencies, which should make the antenna effectively a ``point source'' that should radiate isotropically. Yet, it does not. This is because the so-called ``point source'' has {\it internal anisotropy} accruing from the fact that the power and phase profiles of the spin waves excited within the nanomagnets are spatially inhomogeneous. 

Can one extract the radiation pattern of just the nanomagnets alone somehow? This is prevented by interference. Let us say that the amplitude of the radiation emitted by the nanomagnets and received at some location in space is {\bf n} and that emitted by the peripherals and received at the same location is {\bf p}. Then the total power received at that location is |{\bf n} + {\bf p}e$^{i \phi}$ |$^2$ = ${\bf n}|^2 + |{\bf p}|^2 + 2 |{\bf n} ||{\bf p}|cos \phi$ where $\phi$ is the phase difference between the two amplitudes. This shows us that the power received from the real sample at any location is {\it not} the arithmetic sum of the powers from the nanomagnets and the peripherals since $\phi$ can be anything between 0$^{\circ}$ and 180$^{\circ}$. Hence subtracting the latter from the former does not yield the power radiated by the nanomagnets alone.

\subsubsection{Beam steering}

\begin{figure*}[!hbt]
\centering
\includegraphics[width=0.8\textwidth]{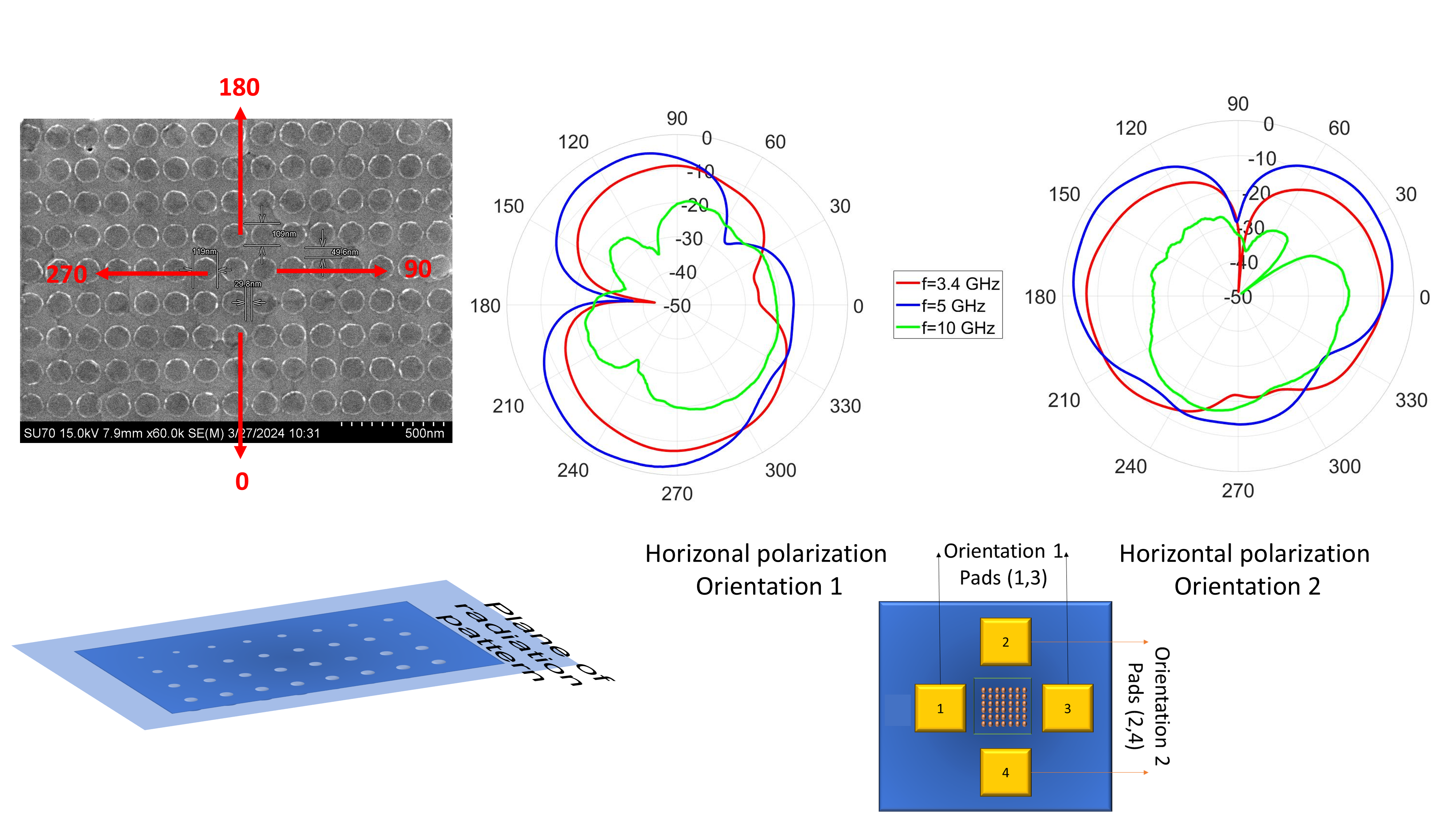}
\vspace{1cm}
\includegraphics[width=0.8\textwidth]{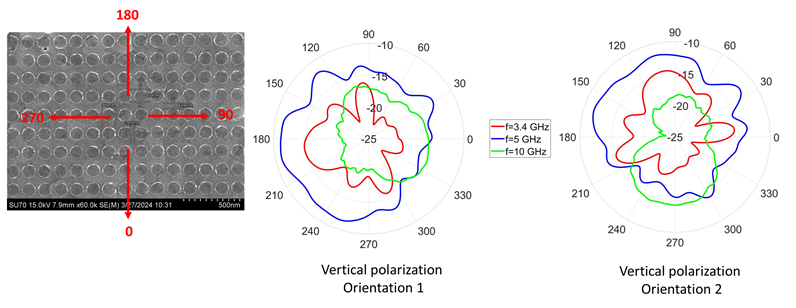}
\caption{\small The radiation pattern of the real sample at three  different frequencies (3.4, 5 and 10 GHz) in the plane of the nanomagnets. The patterns are shown for both (a) horizontal and (b) vertical polarizations of the emitted beam. The radiation pattern for the control sample was already shown in Fig. \ref{fig:control} for horizontal polarization. For vertical polarization, there are similar differences between the radiation patterns of the real and control samples and hence the radiation pattern for the control sample is not shown. The horizontally polarized component is much stronger than the vertically polarized component in this plane. Note the very significant differences in the radiation patterns between orientation 1 (charge current flows between contact pads 1 and 3) and orientation 2 (charge current flows between contact pads 2 and 4) for both polarizations. This difference allows one to change the radiation pattern significantly by changing the direction of current flow, attesting to the beam steering capability. Note that for the two lower frequencies of 3.4 GHz and 5 GHz, {\it the radiation pattern rotates by approximately 90$^{\circ}$ when we rotate the direction of current flow by 90$^{\circ}$}. This is a direct manifestation of beam steering. Reproduced from \cite{TI-antenna} with permission of the American Institute of Physics.}
\label{fig:Pattern1}
\end{figure*}

\begin{figure*}[!hbt]
\centering
\includegraphics[width=0.8\textwidth]{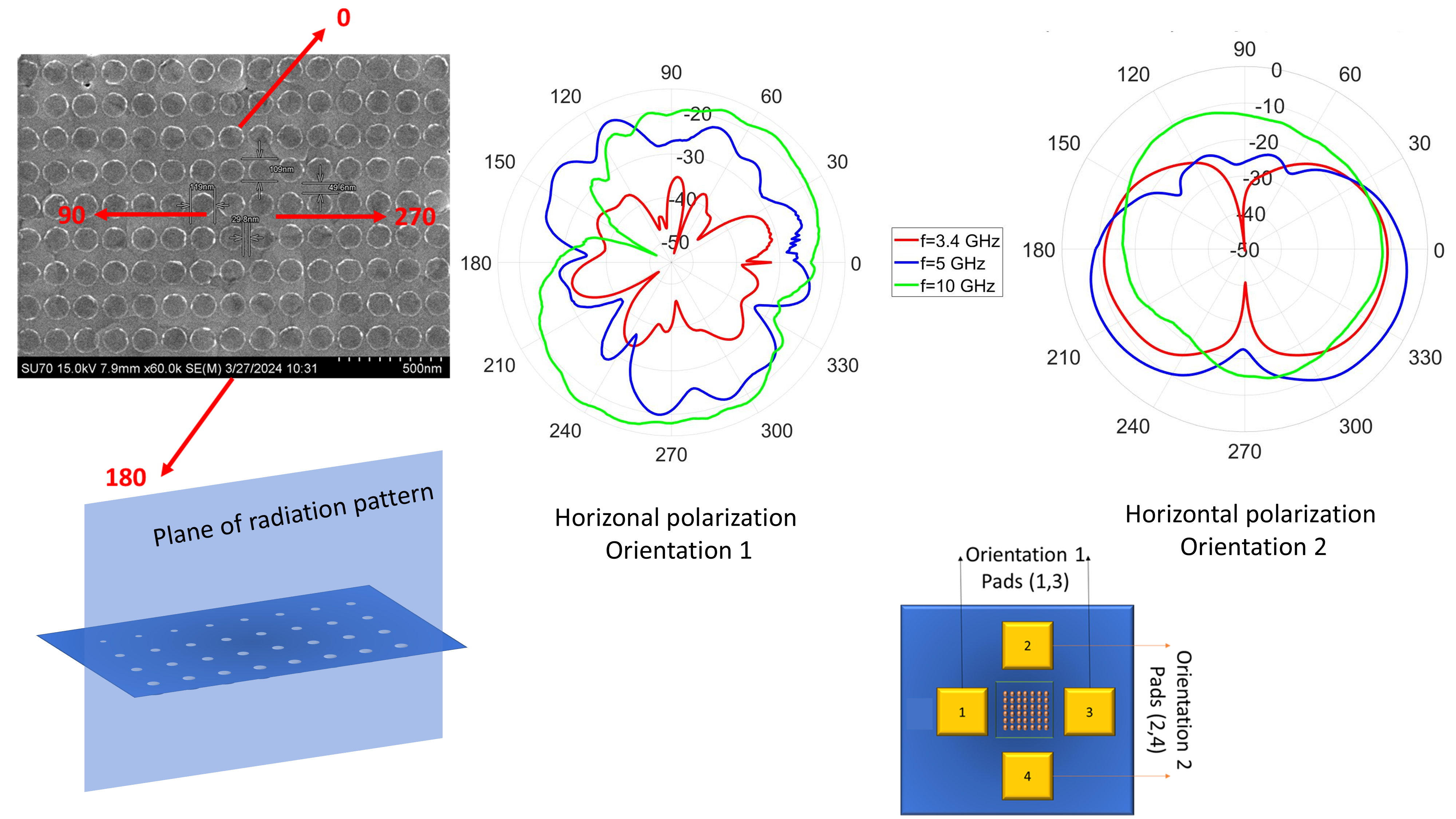}
\includegraphics[width=0.8\textwidth]{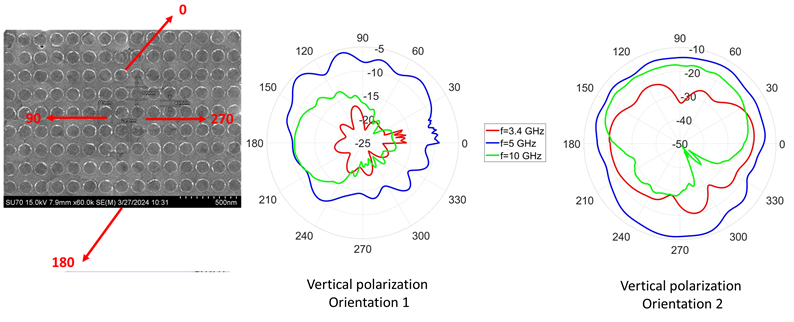}
\caption{\small The radiation pattern of the real sample at three different frequencies of 3.4, 5 and 10 GHz in a plane transverse to that of the nanomagnets. The radiation pattern of the control sample is again quite different from that of the real sample in some directions (as in Fig. \ref{fig:Pattern1}) and is not shown here to avoid repetition. The patterns are shown for both (a) horizontal and (b) vertical polarizations.  Again, note the significant differences in the radiation patterns for orientation 1 (charge current flows between contact pads 1 and 3) and orientation 2 (charge current flows between contact pads 2 and 4). Reproduced from \cite{TI-antenna} with permission of the American Institute of Physics.}
\label{fig:Pattern2}
\end{figure*}

\begin{figure*}[!hbt]
\centering
\includegraphics[width=0.8\textwidth]{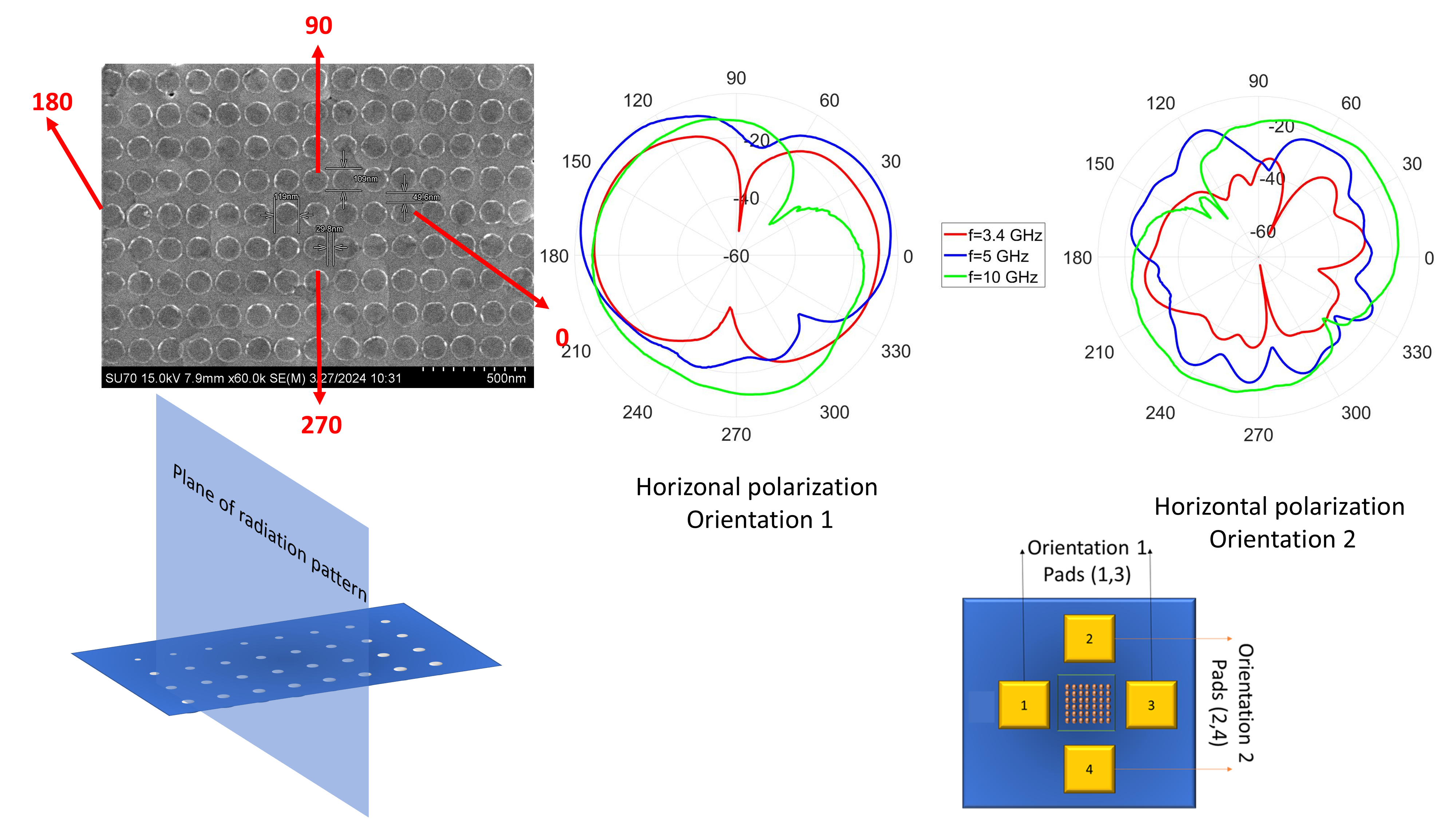}
\vspace{1cm}
\includegraphics[width=0.8\textwidth]{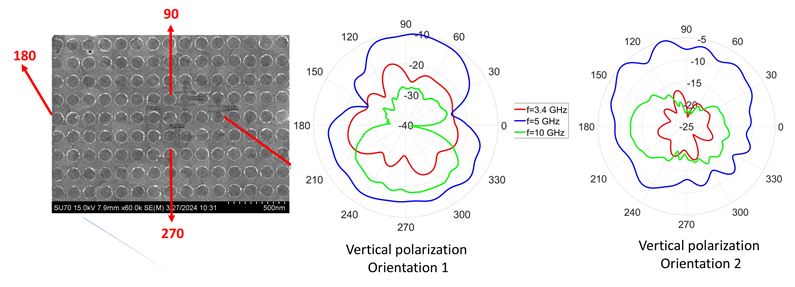}
\caption{\small The radiation pattern of the real sample at frequencies of 3.4, 5 and 10 GHz in the other plane transverse to that of the nanomagnets. Again, the radiation pattern of the control sample is quite different from that of the real sample in some directions (as in Fig. \ref{fig:Pattern1}) and is not shown here to avoid repetition. The patterns are shown for both (a) horizontal and (b) vertical polarizations. Again, note the stark differences in the radiation patterns for orientation 1 (charge current flows between contact pads 1 and 3) and orientation 2 (charge current flows between contact pads 2 and 4). Reproduced from \cite{TI-antenna} with permission of the American Institute of Physics.}
\label{fig:Pattern3}
\end{figure*}

The radiation patterns of the real sample for both horizontal and vertical polarizations of the emitted wave in the plane of the nanomagnets and the two transverse planes were reported in ref. \cite{TI-antenna} and shown in Figs. \ref{fig:Pattern1}, \ref{fig:Pattern2} and \ref{fig:Pattern3} for two mutually perpendicular directions of charge current flow (``orientation 1'' and ``orientation 2''). Perhaps the most interesting observation is that the radiation pattern depends strongly on the direction of charge current flow (note the differences between orientations 1 and 2 in Fig. \ref{fig:Pattern1}). This is not surprising since the charge current's direction determines the axis of the alternating spin polarization within a nanomagnet owing to {\it spin-momentum locking}. Thus, by changing the direction of current flow, we can change the spin polarization axis and hence  the spin wave pattern  in different nanomagnets. Since it is the spin waves that radiate the electromagnetic waves, changing the spin wave patterns will change the radiation pattern, which effectively implements beam steering. Once again, beam steering is accomplished without any phased array and the beam steerer is orders of magnitude smaller than any phased array.

A somewhat confounding observation that one can make in  Fig. \ref{fig:control} is that changing the direction of the current (i.e., changing from orientation 1 to orientation 2) not only changes the radiation pattern of the real sample, but it also changes that of the control sample! The reason for this puzzling observation was explained in Section 6 of the Supporting Information in Ref. \cite{TI-antenna} and not repeated here. Because of this observation, we have to check if the radiation patterns of the real sample and the control sample {\it change by approximately the same amount when we rotate the current direction by the same angle in both samples}. If they do, then we cannot claim beam steering since then we cannot conclude with certainty that the radiation pattern of the nanomagnets depends on current direction. Fortunately, that does not happen. They change in different ways, especially at the lowest frequency of 3.4 GHz, which tells us that the radiation pattern of the nanomagnets alone indeed changes if we change the current direction.

In Fig. \ref{fig:Pattern1} we note that at the two lower frequencies of 3.4 and 5 GHz, rotating the direction of the current by 90$^{\circ}$ {\it also rotates the radiation pattern approximately by 90$^{\circ}$ for the horizontal polarization}, which is the dominant polarization in the emitted beam.  This feature is further evidence of beam steering and it is not observed in the control sample in Fig. \ref{fig:control} which tells us that it originates from the nanomagnets. We also do not see this feature so prominently at the highest frequency of 10 GHz in the real sample because at that high frequency, the spin oscillation may not be able to keep up in synchrony with the charge oscillation, thereby weakening beam steering.

The beam steering effect is less pronounced in the radiation patterns in the two planes that are transverse to the plane of the nanomagnets, as can be seen in Figs. \ref{fig:Pattern2} and \ref{fig:Pattern3}. However the radiation intensities in these two planes are weaker than those in the plane of the nanomagnets, which is not surprising since the spin oscillation takes place in the plane of the nanomagnets. This may be the reason why the beam steering effect is relatively muted in the two transverse planes. 

Finally, we need to assess if this ultra-sub-wavelength quantum-enabled antenna has a radiation efficiency that beats the Harrington limit. This was done in Ref. \cite{TI-antenna} for the cases where the radiation pattern is somewhat isotropic. For those cases, it was found that the radiation efficiency beat the Harrington limit (for an antenna of the same size) by more than {\it two orders of magnitude}. Thus, it shares the two remarkable features of the AMMC antenna: (1) it beats the Harrington limit, and (2) it enables beam steering with a single antenna element.

\section{An ultra-sub-wavelength antenna activated by the giant spin Hall effect in a heavy metal}

We conclude the discussion of this family of novel antennas by introducing the last member which is based on the giant spin Hall effect in a heavy metal like Pt, Ta, etc. which have strong spin-orbit interaction. The spin Hall effect \cite{dyakonov,hirsch} is a remarkable phenomenon in spintronics that is used to inject spin currents into  ferromagnets by passing charge currents through an underlying/overlying heavy metal or topological insulator \cite{mellnik,liu}. This results in a spin-orbit torque (SOT) within the ferromagnet which is harnessed to switch its magnetization, thereby making it an efficient mechanism to write bits into non-volatile magnetic memory \cite{song,memory}. The same effect has also been leveraged to implement combinational Boolean logic using magnetic switches \cite{logic}. These are all digital applications. Analog applications of the spin Hall effect are few and far between, with the most notable application being in the spin Hall nano-oscillator (SHNO) \cite{chen} which is used in microwave assisted magnetic recording \cite{MAMR}, neuromorphic computing \cite{chen}, etc.

Until now, the spin Hall effect has not been used for an antenna.
An {\it antenna} is, of course, very distinct from an {\it oscillator} and has a different application space. Only recently there has been a report of a spin Hall effect based nano-antenna \cite{SHNA}. It was made of a two-dimensional array of nanomagnets placed in contact with a heavy-metal (HM) nanostrip through which an ac current is made to flow. The resulting  spin Hall effect in the HM strip causes alternating spin-orbit torque within the nanomagnets which excites confined spin waves in the nanomagnets at the frequency of the current. As usual, these spin waves transfer their energy to electromagnetic waves via magnon-photon coupling to radiate an electromagnetic wave  into the surrounding medium at the  frequency of the ac current, thereby implementing a ``spin Hall nano-antenna'' (SHNA).  The Supporting Information accompanying Ref. \cite{SHNA} provided a classical phenomenological theory of how alternating spin-orbit torque can produce electromagnetic radiation. This theory is a classical theory based on coupled Maxwell's equation and Landau-Lifshitz-Gilbert equation. It provides a phenomenological explanation of this antenna's operating principle.

The SHNA that was demonstrated in ref. \cite{SHNA} is again an extreme sub-wavelength antenna whose dimensions (sub-mm) were more than an order of magnitude smaller than the radiated electromagnetic wavelength at the frequencies tested, which were between 1 and 10 GHz (free space wavelength 3-30 cm), and yet it radiated efficiently, beating the Harrington limit. However, because of the configuration used, this antenna is not capable of beam steering.

Both a transmitting antenna and a receiving antenna were demonstrated in ref. \cite{SHNA} and both are described in the ensuing sections.

\subsection{Transmitting SHNA}

\subsubsection{Sample description}

\begin{figure}[!h]
\centering
\includegraphics[width=0.99\textwidth]{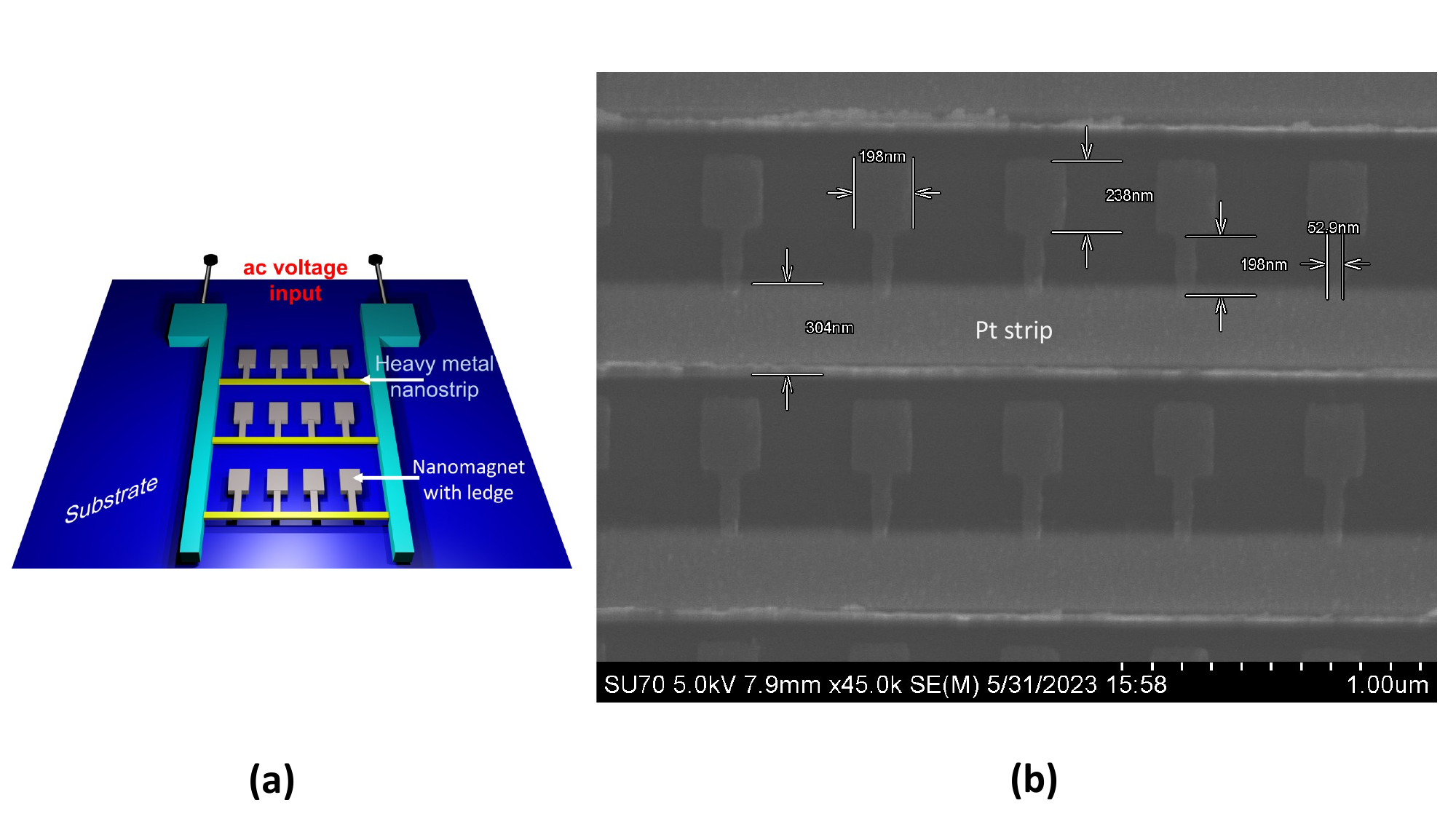}
\caption{(a) Schematic of the spin Hall nano-antenna device. (b) Scanning electron micrograph of the device showing the various structural dimensions). Reproduced from \cite{SHNA} with CC-BY license.}
\label{fig:SHNA}
\end{figure}

The schematic of the SHNA device used in \cite{SHNA} is shown in Fig. \ref{fig:SHNA}(a). It consists of linear periodic arrays of ``ledged'' rectangular magnetostrictive cobalt nanomagnets deposited on a LiNbO$_3$ substrate, with Pt (a heavy metal) nanostrips underlying the ledges as shown in Fig. \ref{fig:SHNA}.  The reason LiNbO$_3$ was used instead of Si and why the nanomagnets were magnetostrictive and had this odd shape with ledges was explained in \cite{SHNA}; the intention was to use this structure  as a {\it dual electromagnetic and acoustic antenna} that radiates an electromagnetic wave into the surrounding medium {\it and} an acoustic wave in the piezoelectric substrate. The acoustic functionality would require a piezoelectric substrate and magnetostrictive nanomagnets. Had we been interested in only an electromagnetic antenna, we could have used a Si substrate and non-magnetostrictive nanomagnets.

The Pt nanostrip was $\sim$300 nm wide and 5 nm thick. There were 3000 linear arrays,  each containing 95 nanomagnets (total of 285,000 nanomagnets), and the ends of the nanostrips in each array were connected to two contact pads as shown in Fig. \ref{fig:SHNA}(a). Alternating charge current was passed between these two pads to generate alternating spin-orbit torques on the nanomagnets, which produced spin waves in the nanomagnets that radiated electromagnetic waves.  Nanomagnet and ledge dimensions, intermagnet separation, etc. are all shown in the scanning electron micrograph in Fig. \ref{fig:SHNA}(b). The inter-nanomagnet separation was large enough that any dipole interaction between neighbors became negligible.

\subsubsection{Transmitting antenna activation}

An alternating charge current pumped into each Pt nanostrip will inject spin currents of alternating spin polarization into the nanomagnets via the spin Hall effect \cite{dyakonov,hirsch}. This causes an alternating spin-orbit torque \cite{ralph} that results in either back-and-forth domain wall motion in the nanomagnets \cite{fert}, or magnetization precession \cite{morrison}, or both. That would excite spin waves within the nanomagnets \cite{morrison,ahsanul}. The spin waves transfer their energy to electromagnetic waves which are  radiated  in the surrounding medium, thereby implementing transmitting antenna action.

\subsubsection{Spin waves in the nanomagnets due to alternating spin Hall effect caused by the alternating current flowing through the Pt nanostrips}

\begin{figure}[!h]
\centering
\includegraphics[width=0.9\textwidth]{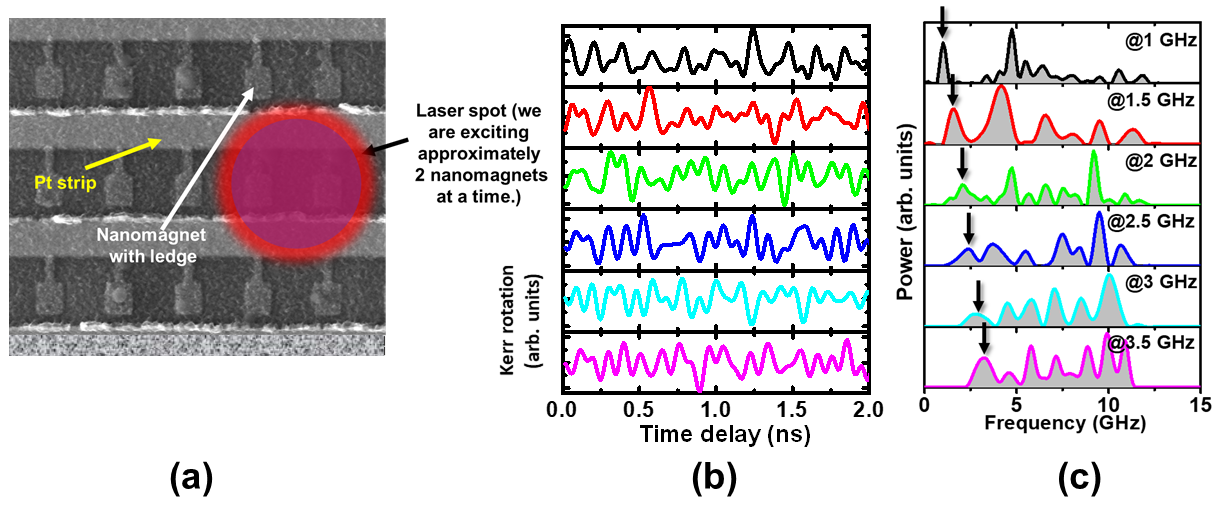}
\caption{(a) The pump and probe laser spots are approximately 1 $\mu$m in diameter and hence cover two nanomagnets at a time. Therefore, the spin waves are always sampled from two nanomagnets. (b) Kerr oscillations in the nanomagnets plotted in the time domain. (c) Fast Fourier transform of the Kerr oscillations showing the peaks in the spin wave spectra. Reproduced from \cite{SHNA} with CC-BY license.}
\label{fig:Kerr}
\end{figure}

To ascertain that pumping an alternating current into the heavy metal nanostrip over(under)lying the ledges of the nanomagnets indeed excites spin waves (magnons) in them, time-resolved magneto-optical Kerr effect (TR-MOKE) measurements were performed in Ref. \cite{SHNA} to confirm that spin waves are generated in the nanomagnets. In these experiments, alternating charge current was pumped into the heavy metal nanostrips at six different frequencies of 1, 1.5, 2, 2.5, 3 and 3.5 GHz with constant input power of 16 dbm. The pump and probe beam laser spots of the TR-MOKE setup overlapped in space and covered only two nanomagnets at a time and hence any spin waves generated were sampled from only two nanomagnets, as shown in Fig. \ref{fig:Kerr}(a). The measured Kerr-oscillations are shown in Fig. \ref{fig:Kerr}(b). Fast Fourier transform of these oscillations revealed the peaks in the spectra of the spin waves excited by the ac current, which are shown in Fig. \ref{fig:Kerr}(c). It is interesting to note that while there is always a peak at the current pumping frequency (as expected), there are also other peaks at higher frequencies which are not necessarily integral multiples of the pumping frequencies.  One might be tempted to guess that they are associated with the ``intrinsic modes'' of this quasi-periodic array which are excited by the alternating current.  However, further investigation showed that the frequencies where these peaks occurred depended on which region of the sample was probed, i.e., they varied from one region of the sample to another, which was not consistent with intrinsic mode behavior. Since  only two nanomagnets were probed at a time, the modes were always probed locally, i.e. within those two nanomagnets. When the pump and probe beam were focused on a {\it different} nanomagnet pair, it was found that the mode at the ac current frequency remained unchanged in frequency, but the other mode frequencies changed.  Thus, the higher frequency modes {\it are different in different regions of the nanomagnet array} and therefore ensemble averaging over the entire array will wash them out. Consequently, one does not see radiation emanating from the antenna sample at their frequencies, but instead sees radiation only at the frequency of the ac current since ensemble averaging does not wash that out.

The origin of the spatially-varying high frequency modes cannot be ascertained with complete certainty, but very likely they are ``vortex modes'' caused by strain pulses generated in the nanomagnet by the laser heating and cooling. The heating and cooling by the pump and probe pulses will cause strain pulses in the nanomagnets because of the unequal thermal expansion coefficients of the nanomagnets and the substrate \cite{sucheta,yahagi,sreya}. It has been shown that such strain pulses spawn vortex modes in magnetostrictive nanomagnets \cite{cui} and that the spectra of these modes depend on the nanomagnet diameter ($D$) and thickness ($t$). Since both $D$ and $t$ vary somewhat across the nanomagnet array, one expects to see variance in the frequencies of the high frequency modes and that is exactly what was seen. Because of this variance, the ensemble average of the high frequency modes across the entire nanomagnet array of 285,000 nanomagnets becomes negligible compared to that of the one occurring at the frequency of the ac current. As a result, no electromagnetic radiation with the frequencies of the spatially-varying high-frequency modes could be observed, but one could only find electromagnetic radiation at the alternating current frequency.

\subsubsection{Radiation patterns of the transmitting SHNA}

\begin{figure}[!hbt]
\centering
\includegraphics[width=0.9\textwidth]{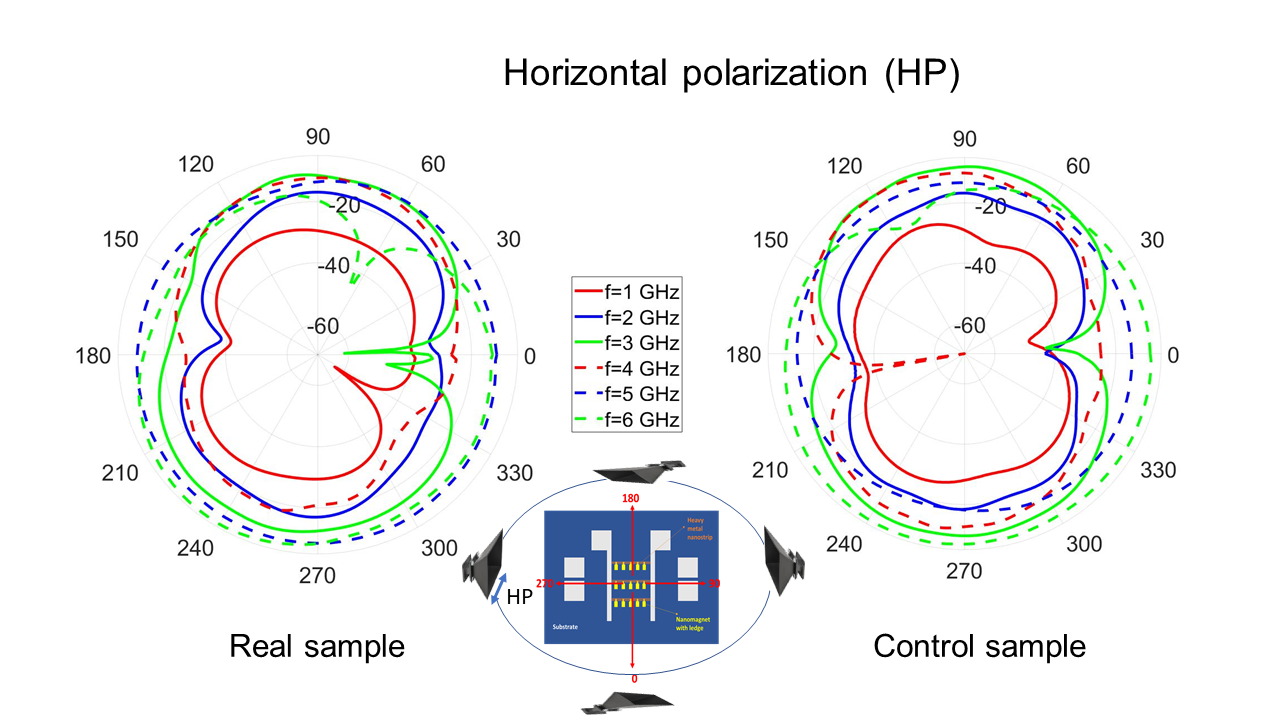}
\includegraphics[width=0.9\textwidth]{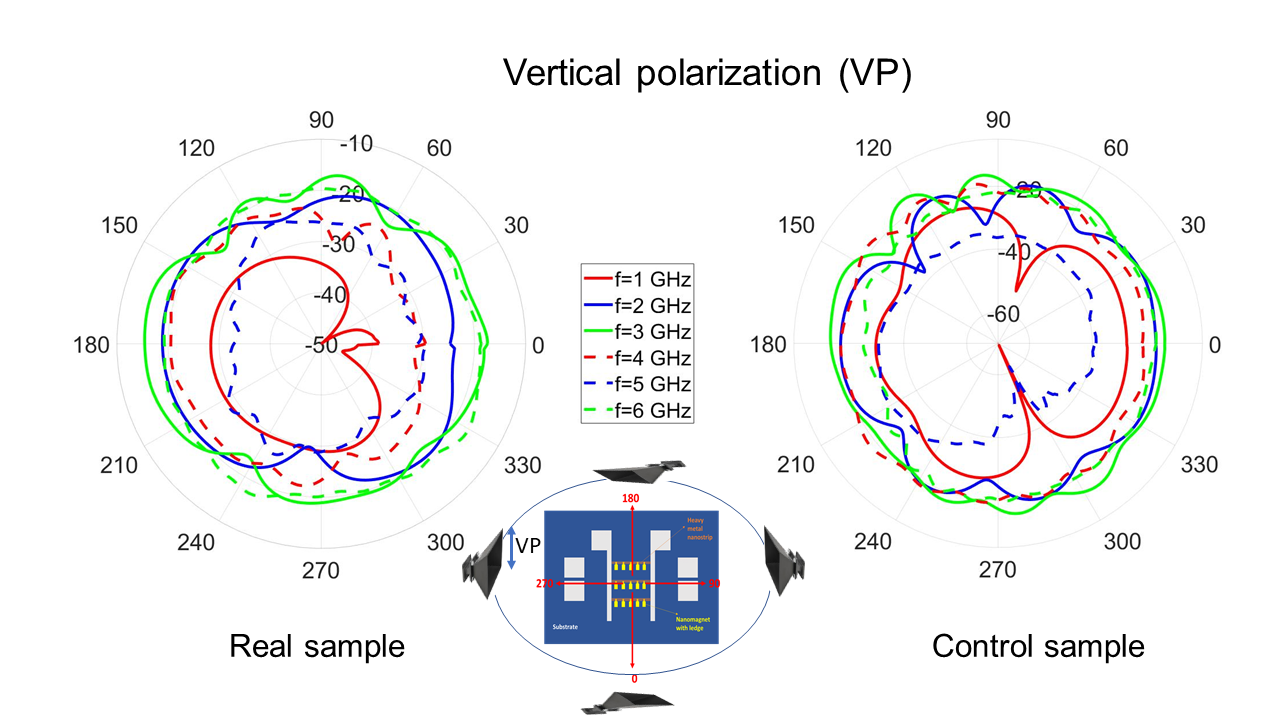}
\caption{The radiation pattern of the SHNA at different frequencies in the plane of the nanomagnets plotted in dbi. The patterns are shown for both the real sample and the control sample, as well as for both horizontal and vertical polarizations. Reproduced from \cite{SHNA} with CC-BY license.}
\label{fig:pattern}
\end{figure}

Ref. \cite{SHNA} reported the radiation patterns of the transmitting SHNA in three different planes -- the plane of the nanomagnets and the two transverse planes. They were measured at frequencies of 1, 2, 3, 4, 5 and 6 GHz (wavelengths 5 - 30 cm). The detector was always placed far enough away from the sample to ensure that the far-field radiation pattern was being measured at all frequencies except 1 GHz. Once again, ``real'' and ``control'' samples were fabricated to ensure that there is radiation from the nanomagnets and hence the radiation  patterns were measured for both the real sample and the control sample. The results are shown in Fig. \ref{fig:pattern} for both horizontal and vertical polarizations in the plane of the nanomagnets at different excitation frequencies. The radiation patterns in the two transverse planes can be found in Section 7 of the Supporting Information of ref. \cite{SHNA}.

Since the radiation pattern of the real sample is very different from that of the control sample, one could conclude with certainty that the nanomagnets were, in fact, radiating.  Once again it was found (as in the case of the nanomagnet/TI antenna) that  the received power from control sample which did not contain the nanomagnets was {\it more} than that from the real sample in some directions. The explanation for this oddity is the same as in the case of the nanomagnet/TI antenna, namely destructive interference. However, once again, in some directions, the power received from the real sample far exceeded that from the control sample. For example, at 1 GHz frequency and for the
vertical polarization, the gain in the 290$^{\circ}$ direction in the plane of the nanomagnets exceeded -30 dbi for the real sample and fell short of -60 dbi for the control sample [see Fig. \ref{fig:pattern}]. This difference of more than 30 dbi (a factor of 1000$\times$) is large enough that it cannot be ascribed to small unavoidable differences between the peripherals (contact pads, etc.) in the two samples. Hence,  {\it the nanomagnets were definitely radiating},
certainly in the 290$^{\circ}$ direction at 1 GHz, and possibly other directions as well at other frequencies.  

Fig. \ref{fig:pattern} also shows that the difference between the radiation from the real sample and the control sample varies quite strongly with direction, which means that the nanomagnets are radiating {\it anisotropically} and in fact the radiation is very ``directional'' since, for example, it is much stronger in the 290$^{\circ}$ direction than in others. Therefore, a strong beam is formed in that direction by the nanomagnets. This anisotropy is surprising since the lateral dimension of the entire nanomagnet array ($\sim$ 160 $\mu$m) is much smaller than the electromagnetic wavelength at all measurement frequencies, which would make the antenna mimic a ``point source'' that is known to radiate omnidirectionally and isotropically.  Yet it does not because of the ``internal anisotropy'' of the source that accrues from the anisotropy of the spin wave patterns forming within the nanomagnets. The spin wave patterns are anisotropic in any case, but here they are even more anisotropic because of the odd shapes of the ledged nanomagnets.

We also note that the radiation  pattern of the control sample (which has no nanomagnets) is anisotropic as well, but this is to be expected. The radiation pattern of almost any sample that is comparable to or larger than the wavelength is naturally anisotropic and the control sample is in that category. Hence, we expect its radiation pattern to be anisotropic. This has no bearing on the fact that the radiation from the nanomagnets is anisotropic and strongly directional.

\begin{figure}[!h]
\centering
\includegraphics[width=0.9\textwidth]{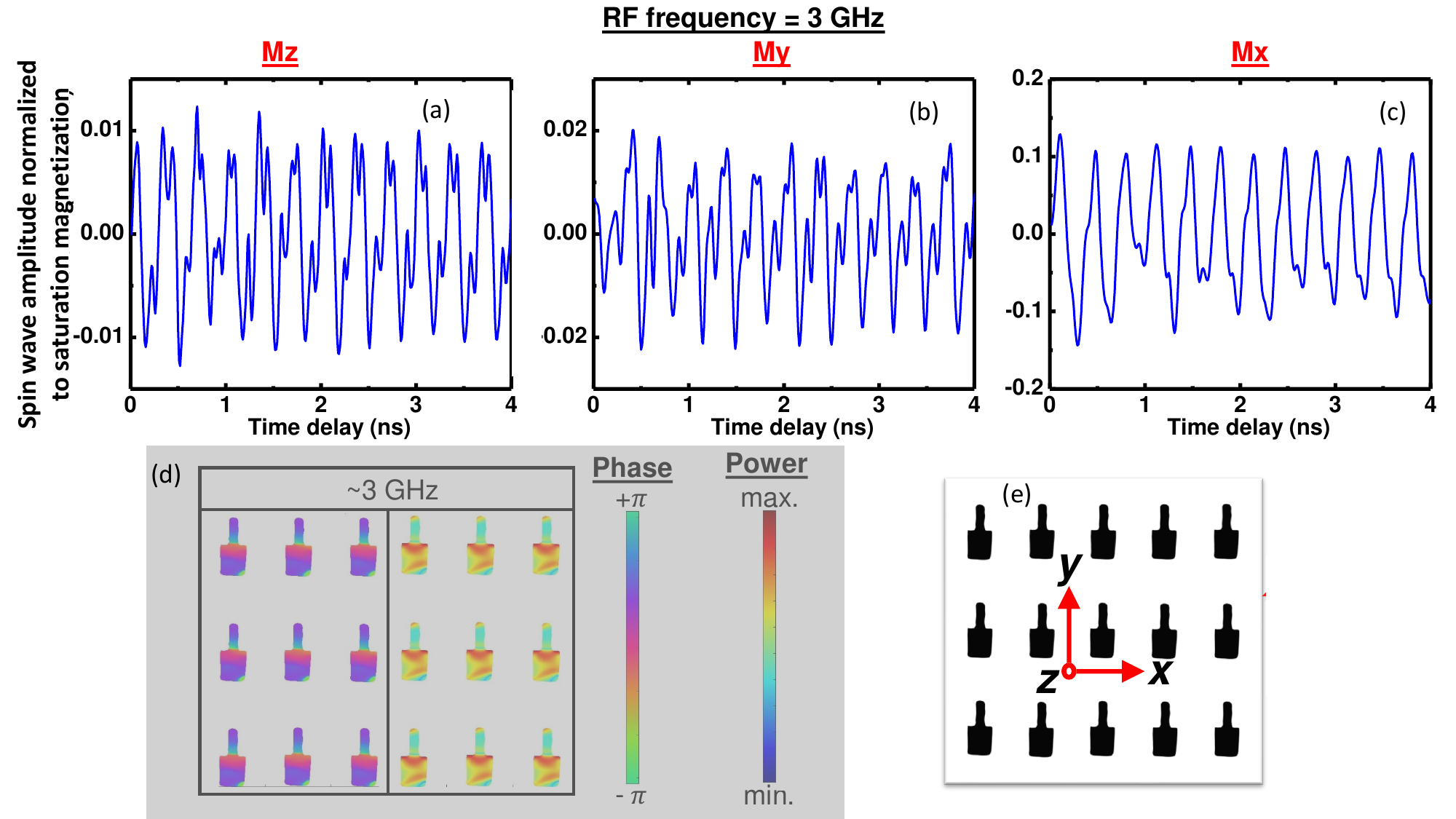}
\caption{Simulated magnetization oscillations (spin waves) in a nanomagnet at an ac current frequency of 3 GHz. (a) z-component (along the thickness); (b) y-component (along the ledge); (c) x-component (perpendicular to the ledge; (d)  phase [left panel] and power [right panel] profiles of the spin waves inside a nanomagnet; (e) designation of the coordinate axes $x$, $y$ and $z$. Reproduced from \cite{SHNA} with CC-BY license.}
\label{fig:oscillations}
\end{figure}

To study the anisotropy of the spin wave patterns excited in the nanomagnets by the spin Hall effect, Ref. \cite{SHNA} used the micromagnetic simulator OOMMF package to find the magnetization components along the three coordinate axes $M_x(t)$, $M_y(t)$ and $M_z(t)$ as a function of time for three different frequencies of the pumping current: 3, 4 and 6 GHz. These plots are shown in Fig. \ref{fig:oscillations} for 3 GHz, which also shows the axes directions. The magnetization oscillations are independent of the initial magnetization states within the nanomagnets. It is interesting to note that the oscillations (and hence the spin waves) have much larger amplitude along  the $x$-direction than along the other two transverse directions ($y$ and $z$). This cannot be explained by the mere fact that the nanomagnets have in-plane anisotropy and hence the in-plane magnetization will be larger than the out-of-plane magnetization. If that were the case, then the x-component should not have been five times larger than the y-component, given that both x- and y-directions are in-plane directions. This unusual feature is observed at all three frequencies of 3, 4 and 6 GHz. Only the 3 GHz results are shown in Fig. \ref{fig:oscillations}, while the other two can be found in Section 8 of the Supporting Information accompanying Ref. \cite{SHNA}. This feature is actually due to the {\it shape} anisotropy of the nanomagnets which have ledges in the y-direction and not in the x-direction. Hence the spin wave amplitude is larger in the x-direction than in the y-direction, even though both are in-plane directions.

Fig. \ref{fig:oscillations} also shows the power and phase profiles of the spin waves excited in the nanomagnet at 3 GHz excitation, calculated by the procedure described in \cite{barman}. Surprisingly, these profiles are quite frequency-dependent and that must be responsible for the {\it frequency dependence} of the radiation pattern (because the spin wave patterns are frequency-dependent).

\subsection{Receiving SHNA}

Having described the transmitting SHNA, we now describe the receiving SHNA which is the electromagnetic reciprocal of the transmitting antenna.

When electromagnetic (EM) radiation is incident on a ferromagnet, it excites spin waves in the latter. This can generally happen in two ways. First, the oscillating magnetic field in the EM radiation can excite spin waves  \cite{hillebrands,ganguly,serga}.
Second, parametric pumping can excite spin waves at half the frequency of the oscillating magnetic field \cite{serga, gurevich,lvov}. An additional consideration comes into play when the ferromagnet is patterned into {\it nanostructures} and arranged in a periodic two-dimensional array, namely a ``magnonic crystal''. In this case, the spin waves that form in the array under different types of excitation produce {\it intrinsic} and {\it extrinsic} modes as discussed earlier \cite{nanoscale}. The intrinsic  mode frequencies are determined by such parameters as the shape, size and material composition of the nanomagnets \cite{suzuki}, as well as the pitch of the array, and may not have any relation to the EM wave frequency. The extrinsic mode frequency, however, is the same as that of the excitation, namely the EM wave and is determined only by the EM wave. Both intrinsic and extrinsic modes are spawned when energy is transferred from the incident EM wave to spin waves within the nanomagnets by photon-magnon coupling which has recently been shown to be quite strong in these systems \cite{salikhov}.

\begin{figure}[!t]
\centering
\includegraphics[width=0.99\textwidth]{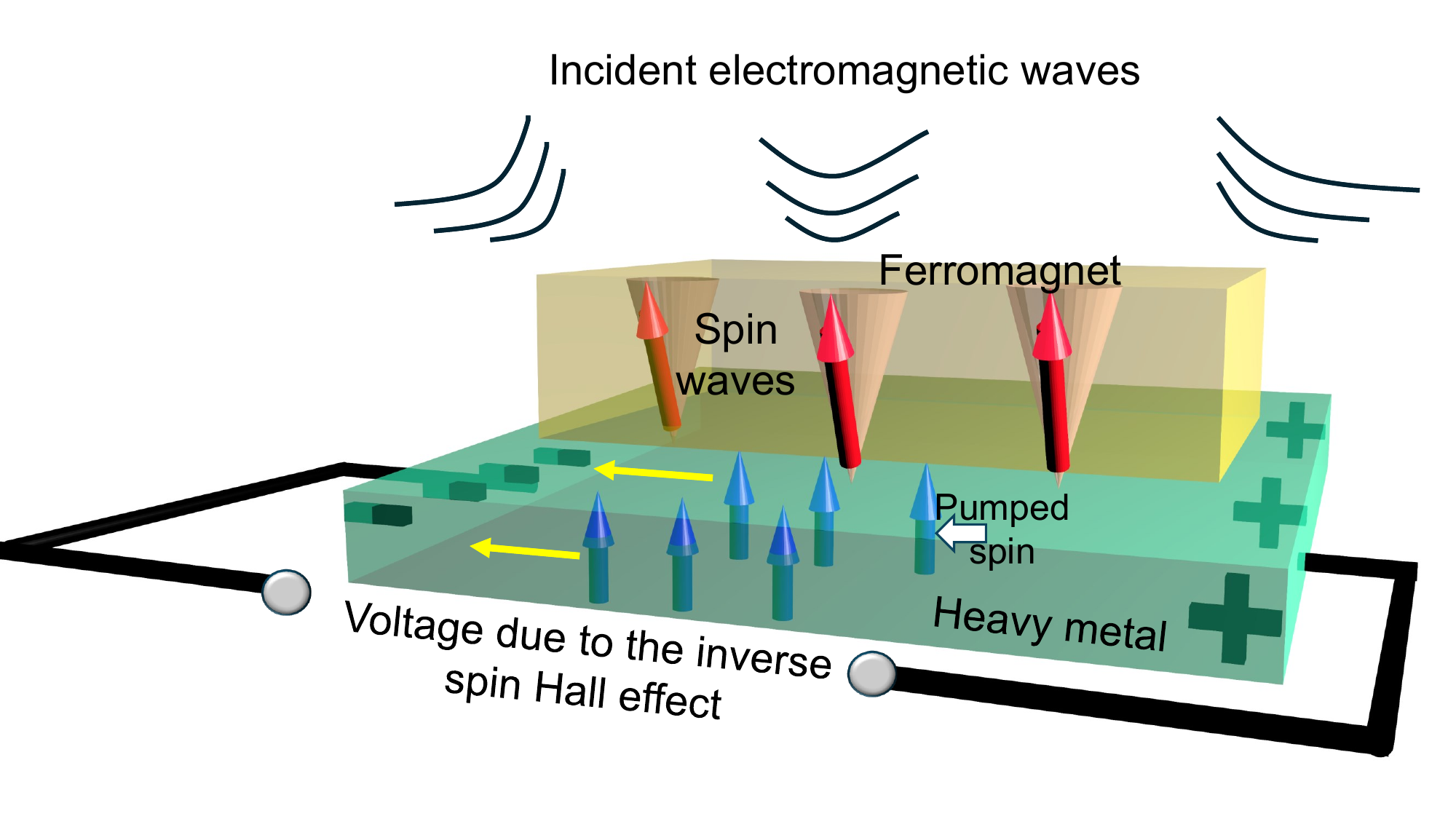}
\caption{Operating principle of the receiving antenna based on ac spin pumping and the ac inverse spin Hall effect. EM radiation excites spin waves in the ferromagnet which pumps spin into the heavy metal layer and that causes an ac voltage to appear across the latter which can be electrically detected to implement a receiving antenna. Reproduced from \cite{SHNA} with CC-BY license.}
\label{fig:principle}
\end{figure}

\begin{figure}[!h]
\vspace{-1in}
\centering
\includegraphics[angle=270,width=5.8in]{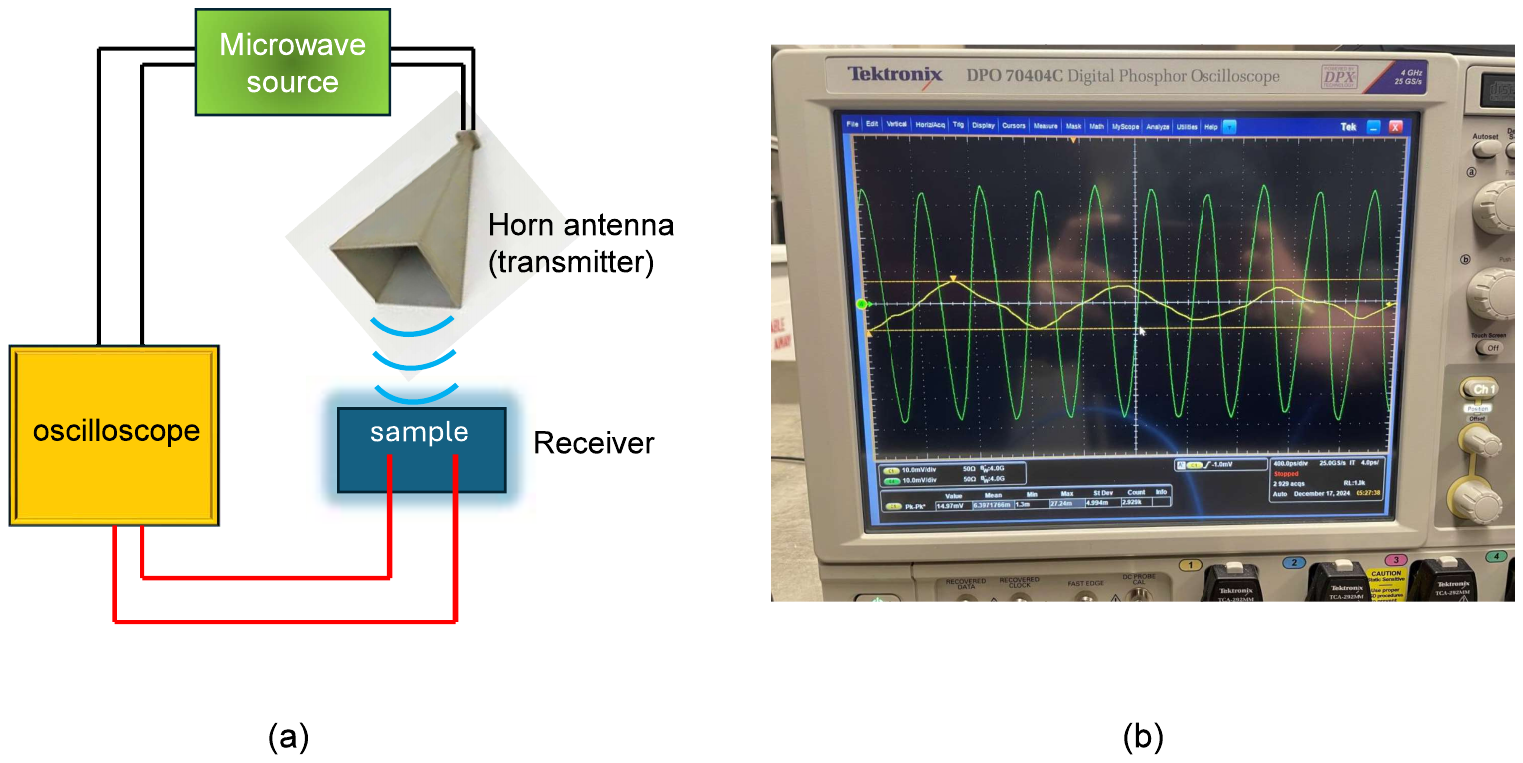}
\vspace{-0.8in}
\caption{(a) Schematic of the experimental set up. (b) Oscilloscope traces for the real sample when the separation between the horn antenna and the sample is 6 in. The green trace is the signal fed to the horn antenna (transmitted signal) and the yellow trace is the output measured between the two contact pads of the real sample (received signal). They are both on the same scale of 10 mV/div. The waveforms and periods are very different since the output has intrinsic and extrinsic modes mixed into it. In this case, the two signals are almost out of phase with each other and the time difference between a peak of the yellow trace and the closest peak of the green trace is 0.13 ns. The ratio of the peak-to-peak amplitudes of the two signals $V_{in}/V_{out}$ is $\sim$4:1. Reproduced from \cite{SHNA} with CC-BY license.}
\label{fig:results}
\end{figure}

If the nanomagnets are in physical contact with a heavy metal that exhibits the spin Hall effect (e.g., Pt), then the EM-excited intrinsic and extrinsic spin waves can {\it pump} spin into the heavy metal \cite{brataas} at their own frequencies and that can cause a polychromatic ac voltage to appear across the heavy metal via the ac inverse spin Hall effect \cite{brataas,silva,youssef,woltersdorf,jiao}. This voltage's frequency components will correspond to frequencies of the intrinsic and extrinsic modes that are excited in the nanomagnets by the EM field. The appearance of this voltage signals the presence of the EM radiation and hence implements a {\it receiving antenna}. The generated voltage will contain frequency components corresponding to the frequencies of the excited intrinsic and extrinsic spin wave modes. The underlying principle of this voltage generation (i.e. the receiver antenna operation) is illustrated in Fig. \ref{fig:principle}. 

\subsubsection{Testing the receiver function}

To demonstrate the receiver function, two samples -- one with the nanomagnets (real sample)  and the other without (control sample) -- were fabricated and tested in ref. \cite{SHNA}. The signals received from the two samples were compared to eliminate spurious effects. Ideally, the control sample should not generate any ac voltage output, but the real sample should.

\begin{figure}[!hbt]
\vspace{-0.8in}
\centering
\includegraphics[angle=270,width=6in]{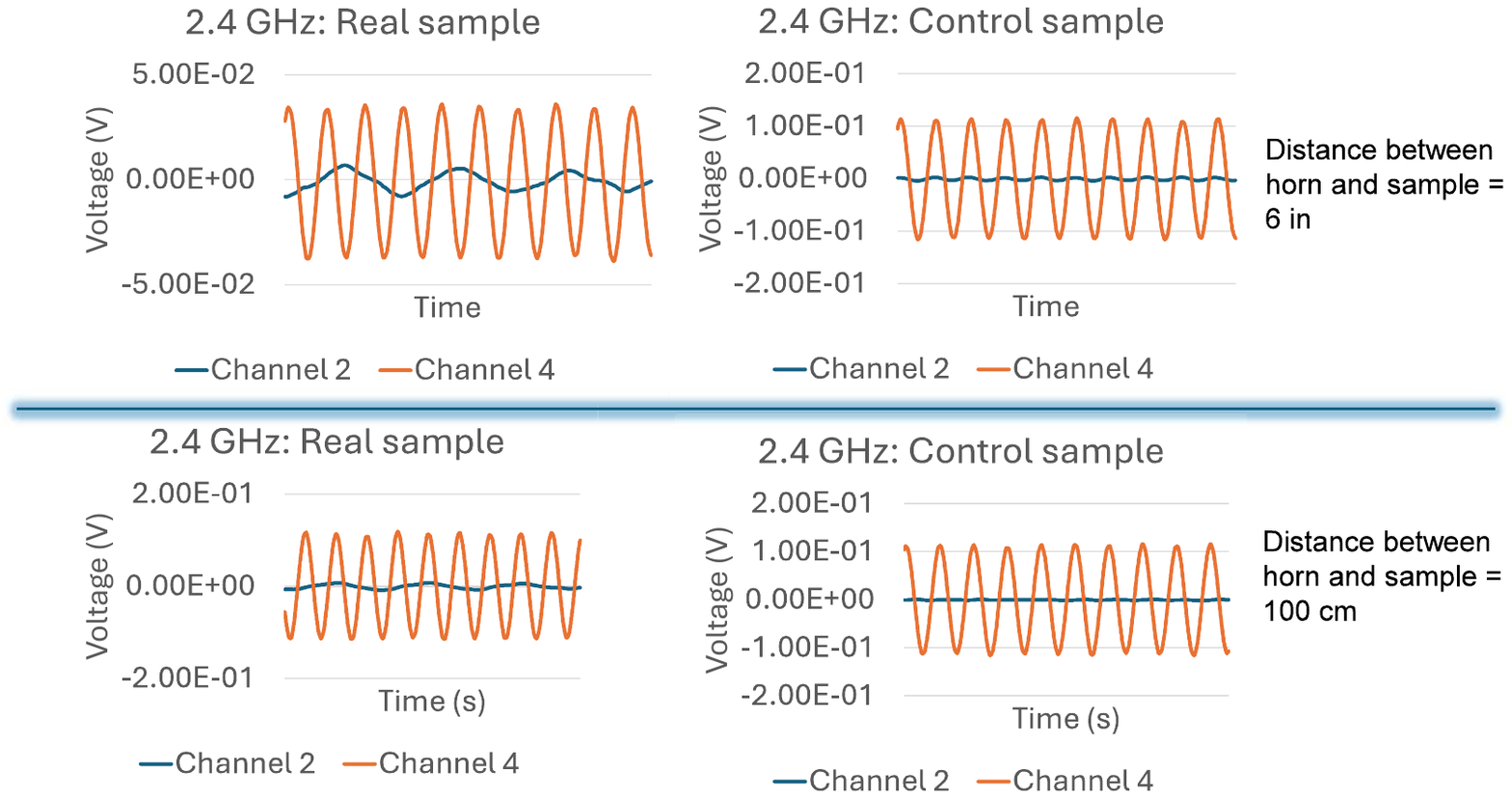}
\vspace{-0.6in}
\caption{Digitized oscilloscope traces of the input signals fed to the horn antenna (in orange) and the output signals produced at the two output contact pads (in blue). The upper panel corresponds to a horn-sample separation of 6 in and the lower panel to 100 cm. The input signal is fed at channel 2 of the oscilloscope and the output signal at channel 4. The left panel corresponds to the real sample and the right panel to the control sample. Reproduced from \cite{SHNA} with CC-BY license.}
\label{fig:results1}
\end{figure}

Both samples were illuminated with a 2.4 GHz and a 1.5 GHz microwave signal transmitted with a horn antenna fed from a microwave source emitting 5 dbm of power as shown in Fig. \ref{fig:results}(a). The two output pads of the sample and the port of the microwave source were connected to two different channels of a microwave frequency digital  oscilloscope. The first channel displayed the waveform of the incident signal emanating from the horn antenna and the second channel displayed the waveform of the received signal measured between the two contact pads. We observed the oscilloscope traces in the two channels at two different separations between the sample and the transmitting horn antenna -- of 6 inches and 100 cm. 
The digital outputs of the oscilloscopes are plotted in  Fig. \ref{fig:results1}  for the two separations at 2.4 GHz excitation. The results for both the real sample and the control sample are shown.
The frequency 2.4 GHz was chosen since it conforms to standard Bluetooth and Wi-Fi.

\subsubsection{Discussion of the receiver function}

Several features are observed in Fig. \ref{fig:results1}. In the {\it control} sample, a small ac voltage appears between the output terminals when the distance between the horn and the sample is 6 inches. Nothing detectable appears when the distance increases to 100 cm (at the oscilloscope scale of 10 mV/div). More importantly, at 6 inches separation, the ac voltage detected at the control sample's output has the {\it same waveform} as the EM signal radiated by the horn antenna and there is virtually {\it no phase shift} between them. Therefore, this received signal in the control sample is most likely due to direct {\it electromagnetic pickup} through the air that has nothing to do with the receiver functionality. Signal from the horn antenna is traveling directly through air to the output contact pads of the control sample. The signal attenuates as it travels through the air and hence it can be detected at 6 inches separation, but not 100 cm (=39 inches) separation.

The story with the {\it real} sample is very different. There are many differences between the signal fed to the horn antenna (transmitted signal) and the signal appearing between the output contact pads (received signal): (1) they do not have the same frequency, (2) they do not have the same waveform and (3) there is a clear phase shift between the two. This eliminates electromagnetic pickup as the source of the output signal. It is the signal produced by transduction of EM waves to spin waves, followed by spin pumping, followed by the ac inverse spin Hall effect that causes the ouput voltage. The existence of this  signal establishes the detector (or receiving antenna) functionality.

\subsubsection{Frequency components of the received signal}

\begin{figure}[!ht]
\centering
\includegraphics[width=6in]{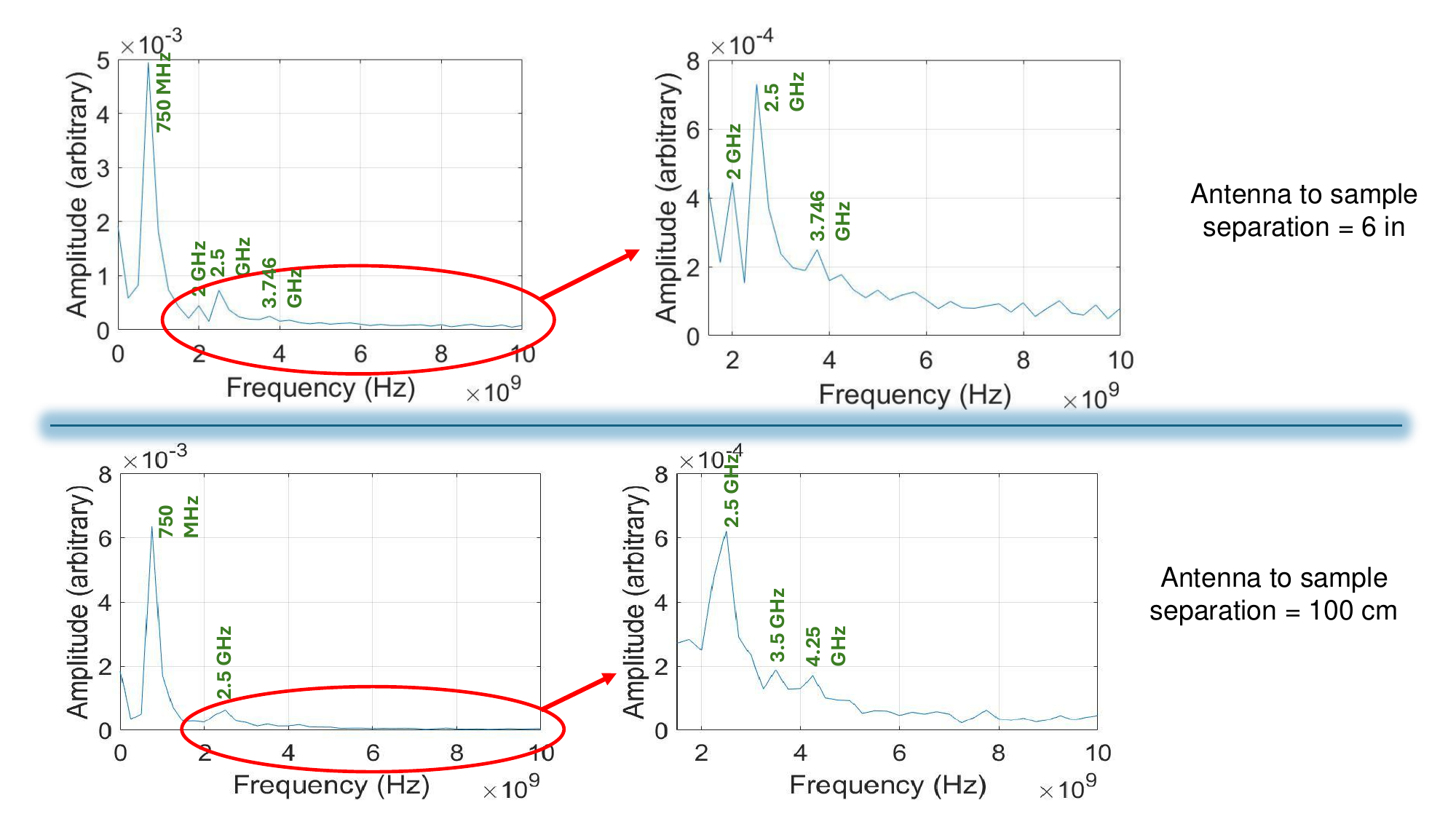}
\caption{Fast Fourier transform (FFT) of the received signal when the EM excitation frequency is 2.4 GHz. The upper panel shows the result when the horn-to-sample separation is 6 in and the lower panel shows the result when the separation is 100 cm. The left figure in either panel is the FFT showing all the peaks, whereas the right figure is a plot of the satellite peaks where the main peak at 750 MHz has been intentionally suppressed. Reproduced from \cite{SHNA} with CC-BY license.}
\label{fig:Fourier}
\end{figure}

Fig. \ref{fig:Fourier} shows the fast Fourier transform of the received signal in  the real sample at 2.4 GHz excitation for the case when the horn-sample separation is 6 in and also for the case when the separation is 100 cm. Note that the dominant peak is at 750 MHz which is {\it not} the EM signal frequency of 2.4 GHz. There is also a satellite peak at 2.5 GHz (close to the 2.4 GHz frequency of the incident radiation). Both peaks show up independent of the horn-to-antenna separation, i.e., both at 6 in and 100 cm. There are other smaller peaks whose frequencies are not separation independent. 

The 750 MHz peak also shows up when the excitation frequency is changed to 1.5 GHz from 2.4 GHz and this was shown in the Supplementary Material accompanying ref. \cite{SHNA}. Very likely this is an {\it intrinsic} spin wave mode in the nanomagnet array which is excited by the EM wave. Its frequency is determined by the size, shape and material composition of the nanomagnets \cite{nanoscale,suzuki} as well as other array parameters, but is independent of extrinsic parameters such as the EM frequency or separation between the transmitter and the receiver. It is excited by energy transfer from photons in the incident electromagnetic wave to magnons in the nanomagnets \cite{salikhov}. 

\subsubsection{Covert communication with SHNA}

The presence of additional frequency components and the resultant waveform distortion in the received signal is actually a boon that enables covert communication with the SHNA. Reassuringly, Fourier transforms of the received signal at either 2.4 GHz incident radiation or 1.5 GHz incident radiation do have frequency components at $\sim$2.4 GHz and 1.5 GHz, respectively. Thus, an intended receiver, who is aware of the transmission frequency can always use a narrow band-pass filter centered at the transmission frequency to filter out the transmitted signal from the received signal. An unintended receiver (eavesdropper), on the other hand, will have no knowledge of the transmission frequency and hence cannot employ this strategy to filter out and receive the transmission. They will be confounded by the additional frequency components. This introduces an element of ``stealth'', whereby a message can be transmitted over a public channel and yet concealed from an eavesdropper. 

\subsubsection{Control sample}

Looking at the results for the {\it control} sample in the right panel of Fig. \ref{fig:results1}, one does not find any measurable signal produced between the contact pads at the horn-sample separation of 100 cm (when both oscilloscope channels have the same amplification of 10 mV/div) and the signal at the separation of 6 in clearly has only one frequency component (or at least a dominant frequency component) at 2.4 GHz which is the signal frequency. There is no frequency component at 750 MHz or 2.5 GHz since the control does not have any nanomagnet. The signal produced between the pads in the control sample is almost surely due to electromagnetic pick up because it is in phase with the input signal (within our measurement tolerance) and has the same frequency. This vindicates the receiver functionality.

\subsubsection{Receiver gain}

Ref. \cite{SHNA} also calculated the receiver gain $G_r$ at 2.4 GHz incident frequency from the formula \cite{Stutzman}
\begin{equation}
\frac{P_r}{P_t} \approx \frac{V_{out}^2}{V_{in}^2}  = G_t G_r {{\lambda^2 }\over{4 \pi R^2}},
    \end{equation}
where $\lambda$ is the free-space wavelength of the incident electromagnetic wave, $R$ is the separation between the antenna and the sample, $P_r$ and $P_t$ are the received and transmitted power, respectively, $V_{out}$ and $V_{in}$ are the amplitudes of the waveform in the two oscilloscope channels, and $G_t$ is the gain of the transmitting horn antenna at 2.4 GHz, which was 9.6 dB = 9.12. This yielded
\begin{equation}
    G_r = 0.1277 = -8.9 db.
\end{equation}

For verification, it was checked if the gain from the 100 cm separation data yielded a very different result. In that case $V{in}/V_{out} \approx 25$. Hence the gain was found to be 
\begin{equation}
    G_r = 0.116 = -9.3 db,   
\end{equation}
which is very close to the value for 6 inches separation showing that the gain is roughly separation-independent at 2.4 GHz incident frequency (as it should be), and is approximately -9 db.

There are theoretical limits on the transmitting gains of conventional antennas that radiate via classical fluctuating electrical dipoles. Although they vary slightly depending on the exact type of the antenna, it is generally of the order \cite{Harrington,skrivervik}
\begin{equation}
    G_t^{max} = A/(2 \pi \lambda )^2 + \sqrt{A}/(\pi \lambda) ,
    \label{limit}
\end{equation}
where $A$ is the antenna area and $\lambda$ is the free space radiated wavelength. Because of the principle of reciprocity, we would expect the same limit to apply to the receiving gain. In our case, this limit turns out to be about 3.22$\times$10$^{-5}$ which is $\sim$ -45 db. 
The gain measured in ref. \cite{SHNA} was 36 db larger, i.e.,  about 4,000 times larger than the theoretical limit for conventional antennas! However, this is not surprising for unconventional antennas that operate by unconventional means.

\section{Conclusion}
This review described a new genre of {\it quantum-enabled} extreme sub-wavelength spintronic transmitting antennas with remarkable features: (1) radiation efficiencies that exceed those of conventional antennas of the same size by several orders of magnitude, (2) beam steering with a single antenna element much smaller than the free space electromagnetic wavelength, and (3) very high radiation efficiencies (exceeding 50\%) at specific frequencies due to resonant transfer of energy from phonons to photons.

The underlying principles are very unconventional, such as tripartite phonon-magnon-photon coupling in artificial multiferroic magnonic crystals, spin injection from the spin-polarized spin-momentum locked surface states of a topological insulator into nanomagnets to spawn spin waves and enable beam steering, and also spin injection from a heavy metal into nanomagnets due to the giant spin Hall effect as well as spin pumping and the inverse spin Hall effect to implement the receiver functionality. These are unprecedented feats and can write a new chapter in fundamental antenna science.

With respect to the AMMC antenna, we point out that these are {\it not} ordinary magneto-elastic antennas and the radiation is {\it not} due to magnetic dipoles which is a very classical effect. In fact, the supplementary material accompanying ref. \cite{SHNA} showed that magnetic dipole radiation would have been $\sim$7 orders of magnitude weaker than what these antennas were shown to radiate. This was later confirmed in \cite{fu1}.  

The applications of these antennas are in all manner of embedded applications that demand extreme antenna miniaturization (orders of magnitude smaller than the free space wavelength) without sacrificing radiation efficiencies and gain. That requires overcoming the Harrington limit \cite{Harrington} by several orders of magnitude, which these antennas have been able to do. The ability to steer a beam with a single ultra-sub-wavelength antenna is another capability which is unprecedented and may find applications in stealth devices and aggressively miniaturized MIMO antennas.

\section*{Data Availability} 

All data generated are already available in the paper.

\section*{Acknowledgements}
The author acknowledges very fruitful collaborations with Prof. Anjan Barman of the S. N. Bose National Center for Basic Sciences, Kolkata, India on carrying out time-resolved magneto-optical Kerr effect microscopy to characterize spin waves in the nanomagnets and Prof. Erdem Topsakal of the author's department and university for supervising the students carrying out anechoic chamber measurements of radiation spectra, scattering parameters and radiation patterns. Two of the authors past Ph.D. students collected most of the data reported here. They are Dr. Justine Drobitch and Dr. Raisa Fabiha.

\section*{Competing Interests}
The author declares no competing interest.

\end{document}